\documentclass[journal]{IEEEtran}

\usepackage{cite}
\usepackage{amsmath,amsthm,amssymb,amsfonts} 
\usepackage{algorithm}
\usepackage{algorithmic}
\usepackage{enumerate}
\usepackage{graphicx}
\usepackage{epstopdf}
\usepackage{url}
\usepackage{bm}
\usepackage{color}

\newtheorem{theorem}{Theorem}

\newtheorem{lemma}{Lemma}
\newtheorem{corollary}{Corollary}

\newtheorem{assumption}{Assumption}

\newcommand{\tabincell}[2]{
	\begin{tabular}{@{}#1@{}}
		
		#2
	\end{tabular}
}

\IEEEoverridecommandlockouts

\begin{document}

\title{Hybrid Duplex Switching in Heterogeneous Networks}

\author{
	\IEEEauthorblockN{Weijun Tang, Suili Feng, Yuan Liu, and Yuehua Ding}
	\thanks{ 
	W. Tang, S. Feng, Y. Liu, and Y. Ding are with School of Electronic and Information Engineering, South China University of Technology, Guangzhou 510641, China (e-mail: tang.weijun@mail.scut.edu.cn, \{fengsl, eeyliu, eeyhding\}@scut.edu.cn).}
}
\maketitle

\begin{abstract}
In this paper, a novel hybrid-duplex scheme based on received power is proposed for heterogeneous networks (HetNets).
In the proposed scheme, the duplex mode (half- or full-duplex) of each user is switchable according to the received power from its serving base station (BS).
The signal-to-interference-plus-noise-ratio (SINR) and spectral efficiency are analyzed for both downlink and uplink channels by using the tools of \emph{inhomogeneous} Poisson point process.
Furthermore, determining power threshold for duplex mode switching is investigated for sum rate maximization, which is formulated as a nonlinear integer programming problem and a greedy algorithm is proposed to solve this problem.
The theoretical analysis and the proposed algorithm are evaluated by numerical simulations.
Simulation results show that the proposed hybrid-duplex scheme outperforms the half-duplex or full-duplex HetNet schemes.

\end{abstract}


\IEEEpeerreviewmaketitle

\section{Introduction}
Due to the overwhelming crosstalk between the transmitter and the receiver circuits, time division duplexing (TDD) and frequency division duplexing (FDD) are commonly used in current wireless systems.
As half-duplex techniques TDD/FDD split a common pool of resources into two subsets for forward and reverse links respectively. 
Such orthogonal resource separation in either temporal or spectral domain might restrict the system performance.
To overcome the bottleneck, full-duplex is believed to be a promising technology, which enables simultaneous transmission and reception on the same frequency for a device.
Thus, the system capacity is significantly improved by doubling spectrum efficiency \cite{sabharwal_-band_2014}.
Unfortunately, the crosstalk, known as self-interference (SI), could be billions of times ($100$ dB+) stronger than the desired signal received over the air \cite{hong_applications_2014}, which prevents full-duplex from engineering application.
Thanks to \cite{bharadia_full_2013,teyan_multi_2015}, a great improvement has been made in self-interference cancellation (SIC). 
Especially, in \cite{teyan_multi_2015}, the SIC capacity up to $-122$ dB is achieved over $20$ MHz band.
These progresses bring in-band full-duplex into practice.

Progress in SIC encourages further researches of full-duplex in cellular networks, such as \cite{goyal_analyzing_2013, mohammadi_full-duplex_2015, psomas_outage_2015, goyal_full_2014}.
With perfect SIC, \cite{goyal_analyzing_2013} investigated a network consisting of full-duplex base stations (BSs) and half-duplex users, whose average user rate was studied by using stochastic geometry.
Under the assumption of imperfect SIC, \cite{mohammadi_full-duplex_2015} derived the closed-form expressions for outage probability and achievable sum rate in a single cell consisting of a full-duplex BS and two half-duplex users.
Both full-duplex BSs and users were considered in \cite{psomas_outage_2015} using stochastic geometry and the outage probability was studied.
In \cite{goyal_full_2014}, a greedy algorithm for joint user selection and power allocation was proposed for multi-cell networks. 
It was shown that scheduling policies greatly impact the system performance.
\cite{goyal_analyzing_2013, mohammadi_full-duplex_2015, yao_secure_2016, psomas_outage_2015, goyal_full_2014} concentrated on homogeneous cellular networks.
For heterogeneous networks (HetNets) supporting explosive growth of mobile service, \cite{jo_heterogeneous_2012, tang_joint_2015} analyzed the downlink SINR in $K$-tier HetNets.
The authors of \cite{lee_hybrid_2015} proposed hybrid full/half-duplex HetNets where the BSs operated in either downlink half-duplex mode or bidirectional full-duplex mode under a predefined probability and all users in a cell used the same duplex mode.
However, as pointed out in \cite{blaszczyszyn_clustering_2015}, the point process of uplink interfering users is not a PPP but a Poisson-Voronoi perturbed lattice.
The authors in \cite{singh_joint_2015} proposed an inhomogeneous PPP model to capture the nature of such point process, where the intensity measure functions of interfering users were derived by considering user association.
The inhomogeneous PPP model was also adopted in \cite{martin-vega_analytical_2016} to analyze the uplink performance in HetNets with power control.

To the best of our knowledge, HetNets whose users are free to switch the duplex modes have not been considered yet, which is the motivation of this paper.
The main contributions of this paper are:

\begin{enumerate}
	\item A novel hybrid-duplex scheme based on the received power is proposed.
	This is motivated by our analysis which reveals that, with imperfect SIC, half-duplex outperforms full-duplex if the downlink received power is lower than a certain threshold.
	In our paper, each user is capable of half-duplex or full-duplex operation according to its received signal power.
	However, reference \cite{lee_hybrid_2015} considered that all users in a same cell operated in either half-duplex or full-duplex mode.
	The simulations show that our proposed scheme significantly outperforms the scheme in \cite{lee_hybrid_2015}.
	\item The threshold for duplex mode switching is discussed. 
	The threshold selection problem is formulated as a nonlinear integer programming for sum rate maximization.
	A greedy algorithm is proposed for solving the problem.
	\item Both downlink and uplink analysis are based on inhomogeneous PPP which is more realistic than homogeneous PPP considered in \cite{lee_hybrid_2015}.
	\item Valuable insights are provided for practical designs.
	In particular, full-duplex would degrade SINR performance with RSI and large inter-cell interference.
	The downlink spectral efficiency is doubled by full-duplex with the sacrifices of uplink performance.
\end{enumerate}

The rest of the paper is organized as follows. 
Section \ref{sec_system_model} presents the hybrid-duplex HetNet model. 
Section \ref{sec_analytical_result} provides the analytical results.
The optimization problem and the numerical results are shown in Section \ref{sec_system_optimization} and Section \ref{sec_numerical_results}, respectively.
Section \ref{sec_conclusion} summarizes this paper.

\section{System Model}
\label{sec_system_model}

\begin{figure}[t]
	\centering
	\includegraphics[width=85mm]{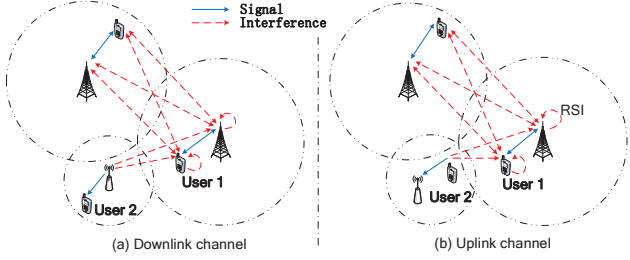}
	\vspace{-0.1cm}
	\caption{A two-tier hybrid-duplex heterogeneous networks.}
	\label{fig_FD_HetNet}
	\vspace{-0.3cm}
\end{figure}

We consider a two-tier network as shown in Fig. \ref{fig_FD_HetNet}.
The BS locations of tier $k$ ($k=1,2$) are distributed as an independent homogeneous Poisson point process (PPP) $\Phi_k$ with density $\lambda_k$.
Without loss of generality, we let the macro cells be tier 1 and the small cells be tier 2. 
The locations of users (denoted by $\mathcal{U}$) follow an independent PPP $\Phi_u$ with density $\lambda_u$.
Every BS of tier $k$ has the same transmit power $\{P_k\}_{k=1,2}$ over the bandwidth $W$, while the transmit power of the users is $P_u$.
Both BSs and users are full-duplex capable. 
Considering imperfect SIC, the residual self-interference (RSI) of the BSs in the two tiers and the users are denoted as $\{RSI_k\}_{k =1,2}$ and $RSI_u$, respectively.
The small-scale fading is assumed to be Rayleigh distribution with unit power.
The path loss is assumed as $l(d) = d^{-\alpha}$, where $d$ is the propagation distance, and $\alpha > 2$ is the path loss exponent.
For simplicity and tractability, shadowing is omitted here. 
Note that shadow fading can be approximately modeled by the randomness of the node locations \cite{lin_optimizing_2015}.
Good approximations are obtained for a realistic assumption (logarithm standard deviation of the shadowing greater than $10$ dB) in many urban scenarios \cite{blaszczyszyn_using_2013}.

\begin{table}[t]
	\renewcommand{\arraystretch}{1.3}
	\caption{Notation Summary}
	\label{table_notation}
	\centering
	\begin{tabular}{|c|c|}
		\hline
		\bfseries Notation & \bfseries Description\\
		\hline
		\hline
		$\Phi_k, \Phi_k^{\text{FD}}, \Phi_k^{\text{HD}}$ & \tabincell{c}{Point process of BSs / full-duplex BSs / \\half-duplex BSs in tier $k$}\\
		\hline
		$\Phi_u, \Phi_{u,k}^{\text{FD}}, \Phi_{u,k}^{\text{HD}}$ & \tabincell{c}{Point process of users / full-duplex users / \\half-duplex users in tier $k$}\\
		\hline
		$\lambda_k; \lambda_u$ & \tabincell{c}{Density of BSs in tier $k$; \\density of mobile users}\\
		\hline
		$P_k; P_u$ & \tabincell{c}{Transmit power of BSs in tier $k$; \\transmit power of users}\\
		\hline
		$RSI_k; RSI_u$ & \tabincell{c}{ RSI of the BSs in tier $k$; RSI of  the users}\\
		\hline
		$\gamma_k$ & \tabincell{c}{Received power threshold of tier $k$}\\
		\hline
		$\alpha; W$ & \tabincell{c}{Path loss exponent; spectrum bandwidth}\\
		\hline
		$D_k$ & \tabincell{c}{Distance from the typical user \\to its closest BS in tier $k$ }\\
		\hline
		$D_k^{\text{FD}}, D_k^{\text{HD}}$ & \tabincell{c}{Transmit distances between the typical user \\and its serving BS in tier $k$}\\
		\hline
		$D_{i,j}$ & \tabincell{c}{Distance between node $i$ and node $j$}\\
		\hline
		$h_{i,j}$ & \tabincell{c}{Small scale fading between node $i$ and node $j$}\\
		\hline
		$\sigma^2$ & \tabincell{c}{Thermal noise power}\\
		\hline
		$\mathcal{A}_k, \mathcal{A}_k^{\text{FD}}, \mathcal{A}_k^{\text{HD}}$ & \tabincell{c}{Association probabilities of the typical user\\ to tier $k$}\\
		\hline
		$I_{t,k}^{i\text{,D}}, I_{t,k}^{i\text{,U}}$ & \tabincell{c}{Cumulative interference on channel $i$ \\from tier $t$ to tier $k$}\\
		\hline
		$\theta_k^{i\text{,FD,D}}, \theta_k^{i\text{,FD,U}}, \theta_k^{i\text{,HD}}$ & \tabincell{c}{SINRs on channel $i$ of a typical user in tier $k$}\\
		\hline
		\tabincell{c}{$\lambda_{k,t}^{\text{BS,FD,D}}, \lambda_{k,t}^{\text{BS,HD,D}}$, \\$\lambda_{k,t}^{\text{BS,FD,U}}, \lambda_{k,t}^{\text{BS,FD,U}}$} &Intensity measure functions of interfering BSs\\
		\hline
		\tabincell{c}{$\lambda_{k,t}^{\text{User,FD,D}}, \lambda_{k,t}^{\text{User,HD,D}}$, \\$\lambda_{k,t}^{\text{User,FD,U}}, \lambda_{k,t}^{\text{User,FD,U}}$} &Intensity measure functions of interfering users\\
		\hline
		$\mathcal{C}_k^{i \text{,FD,D}}, \mathcal{C}_k^{i \text{,FD,U}}, \mathcal{C}_k^{i \text{,HD}}$ & \tabincell{c}{SINR distributions on channel $i$ \\of a typical user in tier $k$}\\
		\hline
		$S_k^{i\text{,FD,D}}, S_k^{i\text{,FD,U}}, S_k^{i\text{,HD}},$ & \tabincell{c}{Spectral efficiency on channel $i$ \\of a typical user in tier $k$}\\
		\hline
	\end{tabular}
\end{table}

We assume that a user is associated with the BS that provides the maximum average downlink received power as in Long Term Evolution (LTE) \cite{3gpp_evolved_2014}.
Let $\{D_k\}$ denote the distance from a typical user to its closest BS in tier $k$. 
The typical user is associated with the BS of tier $k$ if 
\begin{equation}
k = \arg_t\max\{P_t D_t^{-\alpha}, t = 1,2\}.
\end{equation}

After the cell association, the users' duplex modes are scheduled based on the received signal power.
We denote $\{\gamma_k\}_{k = 1,2}$ as the received power thresholds.
If the average downlink received signal power of a macro user is greater than or equal to the threshold of tier $1$, i.e. $P_1 D_1^{- \alpha} \geq \gamma_1$, the user is scheduled as a full-duplex user.
Otherwise, if $P_1 D_1^{- \alpha} < \gamma_1$, the user communicates with its serving BS in half-duplex mode.
For a small cell user, the case is similar except that the received power thresholds is $\gamma_2$.
The reason behind such a duplex mode selection policy is that, with imperfect SIC, using full-duplex mode might not benefit the users with low received signal power.
Hence, the users suffering strong downlink received  power are scheduled in full-duplex mode, otherwise, the users are in half-duplex mode.
More details about the hybrid-duplex mode please refer to Appendix \ref{appendix_policy}.

According to the discussion above, the user set $\mathcal{U}$ can be divided into four disjoint sets:
\begin{equation}
\label{eqn_association_policy}
\begin{cases}
\mathcal{U}_k^{\text{FD}}, \text{if } D_k \leq D_{\bar{k}} \left(\frac{P_k}{P_{\bar{k}}} \right)^{1 / \alpha} \text{ and } D_k \leq \left(\frac{P_k}{\gamma_k}\right)^{1/\alpha}\\
\mathcal{U}_k^{\text{HD}}, \text{if } D_k \leq D_{\bar{k}} \left(\frac{P_k}{P_{\bar{k}}} \right)^{1 / \alpha } \text{ and } D_k > \left(\frac{P_k}{\gamma_k}\right)^{1/\alpha}\\
\end{cases}, 
k = 1,2,
\end{equation}
where $\bar{k} = 2$ if $k = 1$ and $\bar{k} = 1$ if $k = 2$. 
$\mathcal{U}_1^{\text{FD}}$ and $\mathcal{U}_2^{\text{FD}}$ are the full-duplex user sets in macro cells and small cells, respectively.
Similarly, $\mathcal{U}_1^{\text{HD}}$ and $\mathcal{U}_2^{\text{HD}}$ denote the half-duplex user sets in two tiers. 
Furthermore, we define $\mathcal{U}_1 = \mathcal{U}_1^{\text{FD}} \cup \mathcal{U}_1^{\text{HD}}$ and $\mathcal{U}_2 = \mathcal{U}_2^{\text{FD}} \cup \mathcal{U}_2^{\text{HD}}$ as the user sets of macro cells and small cells, respectively.
Obviously, the typical user is in tier $k$ if $D_k \leq D_{\bar{k}} \left( \frac{P_k}{P_{\bar{k}}} \right) ^{1/\alpha}$. 

In this paper, we assume that the users (including full-duplex and half-duplex users) in a cell share the resources equally without intra-cell interference (but inter-cell interference exists due to full spectrum reuse), which can be done by the round-robin scheduling.
Besides, the resources (e.g. time or frequency) in a cell are divided into two orthogonal parts equally:  the uplink channel and the downlink channel.
The two channels are for half-duplex users with uplink and downlink traffic respectively, while the full-duplex users communicate with their serving BSs bidirectionally on both channels.
We call the downlink channel as channel $1$ and the uplink channel as channel $2$ in the rest of this paper.
Further, the infinite backlogged data is assumed, such that the BSs and users always have data to transmit.

We denote $D_k^{\text{FD}}$ and $D_k^{\text{HD}}$ as the transmit distances between the typical user and its serving BS (called the tagged BS) when it is scheduled as a full-duplex user and a half-duplex user in tier $k$, respectively.
The SINRs on channel $1$ of the typical user in tier $k$ can be expressed as 
\begin{align}
\theta_k^{1\text{,FD,D}} &= \frac{P_k h_{b_0, u_{0}} (D_k^{\text{FD}})^{-\alpha}}{RSI_u + \sum\limits_{t = 1,2} I_{t,k}^{1\text{,D}} + \sigma^2},\\
\theta_k^{1\text{,FD,U}} &= \frac{P_u h_{u_{0}, b_0} (D_k^{\text{FD}})^{-\alpha}}{RSI_k + \sum\limits_{t = 1,2} I_{t,k}^{1\text{,U}} + \sigma^2},\\
\theta_k^{1\text{,HD}} &= \frac{P_k h_{b_0, u_{0}} (D_k^{\text{HD}})^{-\alpha}}{\sum\limits_{t=1,2} I_{t,k}^{1\text{,D}} + \sigma^2},
\end{align}
where $\theta_k^{1\text{,FD,D}}$, $\theta_k^{1\text{,FD,U}}$ are the downlink and uplink SINRs if the typical user is a full-duplex user in tier $k$, respectively.
$\theta_k^{1\text{,HD}}$ is the downlink SINR if the typical user is a half-duplex user.
$h_{i,j}\sim \exp(1)$ is the small-scale fading from node $i$ to node $j$. 
$b_0$ is denoted as the tagged BS, and $u_{0}$ is the typical user. 
$\sigma^2$ is the noise power.
$I_{t,k}^{1\text{,D}}$ and $I_{t,k}^{1\text{,U}}$ are the cumulative interference on channel $1$ from tier $t$ ($t = 1, 2$) to the typical users and the tagged BS in tier $k$, respectively.
$I_{t,k}^{1\text{,D}}$ and $I_{t,k}^{1\text{,U}}$ consist of the interference from other BSs and that from the interfering full-duplex users, which can be expressed as:
\begin{align}
\label{eqn_channel_1_downlink_interference}
I_{t,k}^{1\text{,D}} =& \sum\limits_{b_i \in \Phi_t^{\text{FD}} \setminus b_0} P_t h_{b_i, u_{0}} D_{b_i,u_{0}}^{-\alpha} + \sum\limits_{u_i \in \Phi_{u,t}^{\text{FD}} \setminus u_0} P_u h_{u_{i}, u_{0}} D_{u_{i}, u_{0}}^{-\alpha} \nonumber\\
&+ \sum\limits_{b_i \in \Phi_t^{\text{HD}} \setminus b_0} P_t h_{b_i, u_{0}} D_{b_i, u_{0}}^{-\alpha},\\
\label{eqn_channel_1_uplink_interference}
I_{t,k}^{1\text{,U}} =& \sum\limits_{b_i \in \Phi_t^{\text{FD}} \setminus b_0} P_t h_{b_i, b_0} D_{b_i, b_0}^{-\alpha} + \sum\limits_{u_i \in \Phi_{u, t}^{\text{FD}} \setminus u_0} P_u h_{u_{i}, b_0} D_{u_{i}, b_0}^{-\alpha} \nonumber\\
&+ \sum\limits_{b_i \in \Phi_t^{\text{HD}} \setminus b_0} P_t h_{b_i, b_0} D_{b_i, b_0}^{-\alpha},
\end{align}
where $\Phi_t^{\text{FD}}$ denotes the BSs in tier $t$ ($t = 1, 2$) that are scheduling full-duplex users, $\Phi_t^{\text{HD}}$ is denoted as the BSs scheduling half-duplex users.
$u_{b_i}$ is denoted as the interfering user in the BS $b_i$.
$\Phi_{u, t}^{\text{FD}}$ is the point process denoting the locations of full-duplex interfering users in tier $t$.
$D_{i,j}$ is the distance between node $i$ and node $j$.

Similarly, the SINRs on channel $2$ of the typical user in tier $k$ are defined as
\begin{align}
\theta_k^{2\text{,FD,D}} &= \frac{P_k h_{b_0, u_0} (D_k^{\text{FD}})^{-\alpha}}{RSI_u + \sum\limits_{t = 1,2} I_{t,k}^{2\text{,D}} + \sigma^2},\\
\theta_k^{2\text{,FD,U}} &= \frac{P_u h_{u_0, b_0} (D_k^{\text{FD}})^{-\alpha}}{RSI_k + \sum\limits_{t = 1,2} I_{t,k}^{2\text{,U}} + \sigma^2},\\
\theta_k^{2\text{,HD}} &= \frac{P_u h_{u_0, b_0} (D_k^{\text{HD}})^{-\alpha}}{\sum\limits_{t=1,2} I_{t,k}^{2\text{,U}} + \sigma^2},
\end{align}
where $\theta_k^{2\text{,FD,D}}$ and $\theta_k^{2\text{,FD,U}}$ are the full-duplex downlink and uplink SINR in tier $k$, respectively.
$\theta_k^{2\text{,HD}}$ is the half-duplex uplink SINR.
$I_{t,k}^{2\text{,D}}$ and $I_{t,k}^{2\text{,U}}$ are the cumulative interference on channel $2$ from tier $t$ to the typical users and its serving BS in tier $k$, which can be expressed as:
\begin{align}
\label{eqn_channel_2_downlink_interference}
I_{t,k}^{2\text{,D}} = &\sum\limits_{b_i \in \Phi_t^{\text{FD}} \setminus b_0} P_t h_{b_i, u_0} D_{b_i, u_0}^{-\alpha} + \sum\limits_{u_i \in \Phi_{u, t}^{\text{FD}} \setminus u_0} P_u h_{u_{i}, u_0} D_{u_{i}, u_0}^{-\alpha} \nonumber\\
&+ \sum\limits_{u_i \in \Phi_{u, t}^{\text{HD}} \setminus u_0} P_u h_{u_{i}, u_0} D_{u_{i}, u_0}^{-\alpha},\\
\label{eqn_channel_2_uplink_interference}
I_{t,k}^{2\text{,U}} = &\sum\limits_{b_i \in \Phi_t^{\text{FD}} \setminus b_0} P_t h_{b_i, b_0} D_{b_i, b_0}^{-\alpha} + \sum\limits_{u_i \in \Phi_{u, t}^{\text{FD}} \setminus u_0} P_u h_{u_{i}, b_0} D_{u_{i}, b_0}^{-\alpha} \nonumber \\
&+ \sum\limits_{u_i \in \Phi_{u, t}^{\text{HD}} \setminus u_0} P_u h_{u_{i}, b_0} D_{u_{i}, b_0}^{-\alpha},
\end{align}
where $\Phi_{u, t}^{\text{HD}}$ is the point process denoting the locations of half-duplex interfering users in tier $t$.
For ease of reading, the notations are summarized in Table \ref{table_notation}.


\section{Analytical Modeling}
\label{sec_analytical_result}
In this section, we show the analytical performance results of the proposed HetNets. 
First we derive the SINR distributions on channel $i$ ($i \in \{1,2\}$) of the typical user when it is scheduled as a full-duplex or half-duplex user, respectively.
Then we define and derive the spectral efficiency.

\subsection{SINR Distributions}
\label{subsec_sinr_distribution}
The SINR distributions on channel $i$ ($i \in \{1,2\}$) of the typical user are defined as
\begin{align}
\label{eqn_SINR_FD_D_define}
&\mathcal{C}^{i\text{,FD,D}}(T) \nonumber\\
&= \frac{\sum\limits_{k = 1, 2} \mathcal{A}_k^{\text{FD}} \mathcal{C}_k^{i\text{,FD,D}}(T)}{\sum\limits_{k = 1, 2} \mathcal{A}_k^{\text{FD}}} \triangleq \frac{\sum\limits_{k = 1, 2} \mathcal{A}_k^{\text{FD}} \mathbb{P}(\theta_k^{i\text{,FD,D}} > T)}{\sum\limits_{k = 1, 2} \mathcal{A}_k^{\text{FD}}}, \\
\label{eqn_SINR_FD_U_define}
&\mathcal{C}^{i\text{,FD,U}}(T) \nonumber\\
&= \frac{\sum\limits_{k = 1, 2} \mathcal{A}_k^{\text{FD}} \mathcal{C}_k^{i\text{,FD,U}}(T)}{\sum\limits_{k = 1, 2} \mathcal{A}_k^{\text{FD}}} \triangleq \frac{\sum\limits_{k = 1, 2} \mathcal{A}_k^{\text{FD}} \mathbb{P}(\theta_k^{i\text{,FD,U}} > T)}{\sum\limits_{k = 1, 2} \mathcal{A}_k^{\text{FD}}}, \\
\label{eqn_SINR_HD_define}
&\mathcal{C}^{i\text{,HD}}(T) \nonumber\\
&= \frac{\sum\limits_{k = 1, 2} \mathcal{A}_k^{\text{HD}} \mathcal{C}_k^{i\text{,HD}}(T)}{\sum\limits_{k = 1, 2} \mathcal{A}_k^{\text{HD}}} \triangleq \frac{\sum\limits_{k = 1, 2} \mathcal{A}_k^{\text{HD}} \mathbb{P}(\theta_k^{i\text{,HD}} > T)}{\sum\limits_{k = 1, 2} \mathcal{A}_k^{\text{HD}}},
\end{align}
where $\mathcal{C}_k^{i\text{,FD,D}}(T)$ and $\mathcal{C}_k^{i\text{,FD,U}}(T)$ are the downlink and uplink SINR distributions on channel $i$ of the typical user when it is a full-duplex user in tier $k$, respectively.
When the typical user is a half-duplex user in tier $k$, the downlink SINR distribution on channel $1$ is $\mathcal{C}_k^{1\text{,HD}}(T)$ and the uplink SINR distribution on channel $2$ is $\mathcal{C}_k^{2\text{,HD}}(T)$.
$\mathcal{A}_k^{\text{FD}}$ and $\mathcal{A}_k^{\text{HD}}$ are the association probabilities that the typical user is a full/half-duplex user in tier $k$, respectively.
$\mathbb{P}(X)$ denotes the probability that event $X$ happens, and $T$ is the SINR threshold.
Before obtaining the expressions of the SINR distributions, we first derive the association probabilities.
\begin{lemma}
The association probabilities that the typical user is a full-duplex or half-duplex user in tier $k$ are
\begin{align}
\label{eqn_FD_association_probability}
&\mathcal{A}_k^{\text{FD}} = \frac{ \lambda_k P_k^{2/\alpha} \left( 1 - \exp\left( - \pi \lambda_k \delta_k^2 - \pi \lambda_{\bar{k}} \mu_k^2(\delta_k) \right) \right)}{ \lambda_k P_k^{2/\alpha} +  \lambda_{\bar{k}} P_{\bar{k}}^{2/\alpha}} ,\\
\label{eqn_HD_association_probability}
&\mathcal{A}_k^{\text{HD}} = \frac{ \lambda_k P_k^{2/\alpha} \exp\left( - \pi \lambda_k \delta_k^2 - \pi \lambda_{\bar{k}} \mu_k^2(\delta_k) \right)}{ \lambda_k P_k^{2/\alpha} +  \lambda_{\bar{k}} P_{\bar{k}}^{2/\alpha}},
\end{align}
and the probability that a user is associated with tier $k$ is
\begin{equation}
\label{eqn_tier_k_association_probability}
\mathcal{A}_k = \mathcal{A}_k^{\text{FD}} + \mathcal{A}_k^{\text{HD}} = \frac{ \lambda_k P_k^{2/\alpha}}{ \lambda_k P_k^{2/\alpha} +  \lambda_{\bar{k}} P_{\bar{k}}^{2/\alpha}},
\end{equation}
where 
$\delta_k \triangleq \left(\frac{P_k}{\gamma_k}\right)^{1/\alpha}$ and $\mu_k(x) \triangleq x \left(\frac{P_{\bar{k}}}{P_k} \right)^{1/\alpha}, k = 1, 2$.
\end{lemma}

\begin{IEEEproof}
Please see Appendix \ref{appendix_association}.
\end{IEEEproof}

\subsubsection{Downlink SINR Distributions}
\label{subsection_downlink_SINR}
Let us consider the downlink SINR distributions of the typical user.
Note that 
each BS schedules its associated users independently.
Hence, the probability that a BS in tier $t$ schedules a full-duplex user is $\frac{\mathcal{A}_t^{\text{FD}}}{\mathcal{A}_t}$.
According to the property of PPP \cite{baccelli_stochastic_2009}, the BSs ($\Phi_t^{\text{FD}}$) scheduling full-duplex users form a PPP with a density of $\lambda_t^{\text{FD}} \triangleq \frac{\mathcal{A}_t^{\text{FD}}}{\mathcal{A}_t} \lambda_t$.
Similarly, $\Phi_t^{\text{HD}}$ is a PPP with density $\lambda_t^{\text{HD}} \triangleq \frac{\mathcal{A}_t^{\text{HD}}}{\mathcal{A}_t} \lambda_t$.

Due to the scheduling and association criteria, only one user per BS transmits on the same resource as the typical user.
Therefore, $\{\Phi_{u,t}^{\text{FD}}\}$ and $\{\Phi_{u, t}^{\text{HD}}\}$ are not PPPs but Poisson-Voronoi perturbed lattice \cite{blaszczyszyn_clustering_2015}.
In addition, since the users may either be full- or half-duplex according to their received downlink average power, $\{\Phi_t^{\text{FD}}\}$ and $\{\Phi_t^{\text{HD}}\}$ are not PPPs, either.
Moreover, there exists correlation between the typical user/tagged BS and the interfering nodes \cite{singh_joint_2015, martin-vega_analytical_2016}. 
Therefore, we adopt the similar approximation in \cite{singh_joint_2015} to characterize the processes as inhomogeneous PPPs.

\begin{assumption}
	\label{assumption_intensity_downlink}
	When considering the downlink performance in tier $k$, conditioned on the distance $r$ between the tagged BS and the typical user, the point processes are assumed to be Poisson with intensity measure functions as 
	\begin{align}
	\label{eqn_intensity_BS_FD_D}
	&\Phi_t^{\text{FD}}: \lambda_{k, t}^{\text{BS,FD,D}}(y) = \lambda_t^{\text{FD}}\left(1 - \exp(-\lambda_u \pi \zeta^2)\right),\\
	\label{eqn_intensity_BS_HD_D}
	&\Phi_t^{\text{HD}}: \lambda_{k, t}^{\text{BS,HD,D}}(y) = 
	\begin{cases}
	0, \quad \zeta \leq \delta_t,\\
	\lambda_t^{\text{HD}}(1 - \exp(-\lambda_u \pi (\zeta^2 - \delta_t^2))), \zeta > \delta_t,
	\end{cases}
	\end{align}
	\begin{align}	
	\label{eqn_intensity_User_FD_D}
	\Phi_{u, t}^{\text{FD}}: &\lambda_{k, t}^{\text{User,FD,D}}(y) \nonumber\\
	&= \lambda_t^{\text{FD}}\left(1 - \exp\left(-\pi (r + y)^2 \lambda_t (\frac{P_t}{P_k})^{2 / \alpha}\right)\right),\\
	\label{eqn_intensity_User_HD_D}
	\Phi_{u, t}^{\text{HD}}: &\lambda_{k, t}^{\text{User,HD,D}}(y) \nonumber\\
	&= \lambda_t^{\text{HD}}\left(1 - \exp\left(-\pi (r + y)^2 \lambda_t (\frac{P_t}{P_k})^{2 / \alpha}\right)\right),
	\end{align}
	where $y$ is the distance to the typical user, and
	\begin{equation}
	\zeta = \frac{r + y}{\max\{(\frac{P_k}{P_t})^{1/\alpha} - 1, 0\}}.
	\end{equation}
\end{assumption}
More details about these point processes please refer to Appendix \ref{appendix_intensity}.

\begin{theorem}
	\label{theorem_downlink_SINR_distribution}
	The downlink SINR distributions $\mathcal{C}_k^{1\text{,FD,D}}$, $\mathcal{C}_k^{1\text{,HD}}$ and $\mathcal{C}_k^{2\text{FD,D}}$ are \eqref{eqn_channel_1_SINR_FD_D}, \eqref{eqn_channel_1_SINR_HD} and \eqref{eqn_channel_2_SINR_FD_D} on the top of page $5$, respectively,
	\begin{figure*}[t]
		\normalsize
		\begin{align}
		\label{eqn_channel_1_SINR_FD_D}
		\mathcal{C}_k^{1\text{,FD,D}}(T) &= \frac{2 \pi \lambda_k}{\mathcal{A}_k^{\text{FD}}} \int\limits_{0}^{\delta_k} \exp\left(-\frac{T}{SSINR_{k,u}(r)} \right) \mathcal{F}(r) \prod\limits_{t = k, \bar{k}} \mathcal{L}_{\Phi_t^{\text{FD}}}(r,T) \mathcal{L}_{\Phi_{u,t}^{\text{FD}}}(r, T) \mathcal{L}_{\Phi_t^{\text{HD}}}(r, T) dr,\\
		\label{eqn_channel_1_SINR_HD}
		\mathcal{C}_k^{1\text{,HD}}(T) &= \frac{2 \pi \lambda_k}{\mathcal{A}_k^{\text{HD}}} \int\limits_{\delta_t}^{\infty} \exp\left(-\frac{T}{SNR_{k}(r)} \right) \mathcal{F}(r) \prod\limits_{t = k, \bar{k}} \mathcal{L}_{\Phi_t^{\text{FD}}}(r,T) \mathcal{L}_{\Phi_{u,t}^{\text{FD}}}(r, T) \mathcal{L}_{\Phi_t^{\text{HD}}}(r, T) dr,\\
		\label{eqn_channel_2_SINR_FD_D}
		\mathcal{C}_k^{2\text{,FD,D}}(T) &= \frac{2 \pi \lambda_k}{\mathcal{A}_k^{\text{FD}}} \int\limits_{0}^{\delta_t} \exp\left(-\frac{T}{SSINR_{k,u}(r)} \right) \mathcal{F}(r) \prod\limits_{t = k, \bar{k}} \mathcal{L}_{\Phi_t^{\text{FD}}}(r,T) \mathcal{L}_{\Phi_{u,t}^{\text{FD}}}(r, T) \mathcal{L}_{\Phi_{u,t}^{\text{HD}}}(r, T) dr,\\
		\label{eqn_function_F}
		\mathcal{F}(r) &= r \exp\left(-\pi \lambda_k r^2 - \pi \lambda_{\bar{k}}(\mu_k(r))^2 \right),\\
		\label{eqn_Laplace_BS_FD_D}
		\mathcal{L}_{\Phi_t^{\text{FD}}}(r, T) &= \exp\left(-2 \pi \int_{\Delta_{k,t}(r)}^{\infty} \lambda_{k,t}^{\text{BS,FD,D}}(y) y \left(1 - \mathcal{L}_h\left(\frac{P_t T r^{\alpha}}{P_k y^{\alpha}  } \right)\right) dy \right), \\
		\label{eqn_Laplace_BS_HD_D}
		\mathcal{L}_{\Phi_t^{\text{HD}}}(r, T) &= \exp\left(-2 \pi \int_{\Delta_{k,t}(r)}^{\infty} \lambda_{k,t}^{\text{BS,HD,D}}(y) y \left(1 - \mathcal{L}_h\left(\frac{P_t T r^{\alpha}}{P_k y^{\alpha}} \right)\right) dy \right), \\
		\label{eqn_Laplace_User_FD_D}
		\mathcal{L}_{\Phi_{u,t}^{\text{FD}}}(r, T) &= \exp\left(-2 \pi \int_{\max\{r (\frac{P_t}{P_k})^{1 / \alpha} - \delta_t, 0\}}^{\infty} \lambda_{k,t}^{\text{User,FD,D}}(y) y \left(1 - \mathcal{L}_h\left(\frac{P_u T r^{\alpha}}{P_k y^{\alpha}}\right)\right) dy \right),\\
		\label{eqn_Laplace_User_HD_D}
		\mathcal{L}_{\Phi_{u,t}^{\text{HD}}}(r,T) &= \exp\left(-2 \pi \int_{\max\{\delta_t (\frac{P_k}{P_t})^{1 / \alpha} - r, 0\}}^{\infty} \lambda_{k,t}^{\text{User,HD,D}}(y) y \left(1 - \mathcal{L}_h\left(\frac{P_u T r^{\alpha}}{P_k y^{\alpha}}\right)\right) dy \right).
		\end{align}
		\hrulefill
		\vspace*{4pt}
	\end{figure*}
	where $\Delta_{k,t}(r) \triangleq r \left(\frac{P_t}{P_k}\right)^{1/\alpha}$,	$\mathcal{L}_h(s) = \frac{1}{1+s}$ is the Laplace transform of $h \sim \exp(1)$.
	$SSINR_{t,r}(d) \triangleq \frac{P_t d^{-\alpha}}{RSI_r + \sigma^2}$, is the signal-to-self-interference-plus-noise ratio (SSINR), and  $SNR_{t}(d) \triangleq \frac{P_t d^{-\alpha}}{\sigma^2}$ is the signal-to-noise ratio (SNR). 
\end{theorem}
\begin{IEEEproof}
	Please see Appendix \ref{appendix_SINR_D}.
\end{IEEEproof}

\subsubsection{Uplink SINR Distributions}
Adopting a similar approximation to that in the previous subsection, we approximate the point processes of interfering nodes' locations as independent inhomogeneous PPPs.
\begin{assumption}
	\label{assumption_intensity_uplink}
	When considering the uplink performance in tier $k$, conditioned on the distance $r$ between the tagged BS and the typical user, the point processes are assumed to be Poisson with intensity measure functions as
	\begin{align}
	\label{eqn_intensity_BS_FD_U}
	\Phi_t^{\text{FD}}: &\lambda_{k, t}^{\text{BS,FD,U}}(y) = \lambda_t^{\text{FD}}\left(1 - \exp(-\lambda_u \pi \zeta^2)\right),\\
	\label{eqn_intensity_BS_HD_U}
	\Phi_t^{\text{HD}}: &\lambda_{k, t}^{\text{BS,HD,U}}(y) \nonumber\\
	&= 
	\begin{cases}
	0, \quad \zeta \leq \delta_t,\\
	\lambda_t^{\text{HD}}\left(1 - \exp\left(-\lambda_u \pi (\zeta^2 - \delta_t^2)\right)\right), \zeta > \delta_t,
	\end{cases}\\
	\label{eqn_intensity_User_FD_U}
	\Phi_{u, t}^{\text{FD}}: &\lambda_{k, t}^{\text{User,FD,D}}(y) \nonumber\\
	&= \lambda_t^{\text{FD}}\left(1 - \exp\left(-\pi y^2 \lambda_t (\frac{P_t}{P_k})^{2 / \alpha}\right)\right),\\
	\label{eqn_intensity_User_HD_U}
	\Phi_{u, t}^{\text{HD}}: &\lambda_{k, t}^{\text{User,HD,U}}(y) \nonumber\\ 
	&= \lambda_t^{\text{HD}}\left(1 - \exp\left(-\pi y^2 \lambda_t (\frac{P_t}{P_k})^{2 / \alpha}\right)\right),
	\end{align}
	where $y$ is the distance to the tagged BS, and
	\begin{equation}
	\zeta = \frac{y}{\max\{(\frac{P_k}{P_t})^{1/\alpha} - 1, 0\}}.
	\end{equation}
\end{assumption}
The intensity measure functions above are obtained by similar process as that in Appendix \ref{appendix_intensity}.
Due to the limited place, the details are omitted.

\begin{theorem}
	\label{theorem_uplink_SINR_distribution}
	The uplink SINR distributions $\mathcal{C}_k^{1\text{,FD,U}}$, $\mathcal{C}_k^{2\text{,FD,U}}$ and $\mathcal{C}_k^{2\text{,HD}}$ are \eqref{eqn_channel_1_SINR_FD_U}, \eqref{eqn_channel_2_SINR_FD_U}, \eqref{eqn_channel_2_SINR_HD} on the top of page $6$, respectively, where $\mathcal{F}(r)$ is \eqref{eqn_function_F} on page $5$.
	\begin{figure*}[t]
		\normalsize
		\begin{align}
		\label{eqn_channel_1_SINR_FD_U}
		\mathcal{C}_k^{1\text{,FD,U}}(T) &= \frac{2 \pi \lambda_k}{\mathcal{A}_k^{\text{FD}}} \int\limits_0^{\delta_k}  \exp\left(-\frac{T}{SSINR_{u,k}(r)}\right) \mathcal{F}(r) \prod\limits_{t = k , \bar{k}} \mathcal{L}_{\Phi_t^{\text{FD}}}(r, T) \mathcal{L}_{\Phi_{u,t}^{\text{FD}}}(r, T) \mathcal{L}_{\Phi_t^{\text{HD}}}(r, T) dr,\\
		\label{eqn_channel_2_SINR_FD_U}
		\mathcal{C}_k^{2\text{,FD,U}}(T) &= \frac{2 \pi \lambda_k}{\mathcal{A}_k^{\text{FD}}} \int\limits_0^{\delta_k}  \exp\left(-\frac{T}{SSINR_{u,k}(r)}\right) \mathcal{F}(r) \prod\limits_{t = k, \bar{k}} \mathcal{L}_{\Phi_t^{\text{FD}}}(r, T) \mathcal{L}_{\Phi_{u,t}^{\text{FD}}}(r, T) \mathcal{L}_{\Phi_{u,t}^{\text{HD}}}(r, T) dr,\\
		\label{eqn_channel_2_SINR_HD}
		\mathcal{C}_k^{2\text{,HD}}(T) &= \frac{2 \pi \lambda_k}{\mathcal{A}_k^{\text{HD}}} \int\limits_{\delta_k}^{\infty}  \exp\left(-\frac{T}{SNR_u(r)}\right) \mathcal{F}(r) \prod\limits_{t = k, \bar{k}} \mathcal{L}_{\Phi_t^{\text{FD}}}(r, T) \mathcal{L}_{\Phi_{u,t}^{\text{FD}}}(r, T) \mathcal{L}_{\Phi_{u,t}^{\text{HD}}}(r, T) dr,\\
		\label{eqn_Laplace_BS_FD_U}
		\mathcal{L}_{\Phi_t^{\text{FD}}}(r,T) &= \exp\Bigg(-2 \pi p_{k,t}^{I} \int_{0}^{\Delta_k,t(r) + r} \lambda_{k,t}^{\text{BS,FD,U}}(y) y \left(1 - \mathcal{L}_h\left(\frac{P_t T r^{\alpha}}{P_u y^{\alpha}}\right)\right) dy \nonumber\\
		&\qquad\qquad\qquad -2 \pi \int_{\Delta_{k,t}(r) + r}^{\infty} \lambda_{k,t}^{\text{BS,FD,U}}(y) y \left(1 - \mathcal{L}_h\left(\frac{P_t T r^{\alpha}}{P_u y^{\alpha}}\right)\right) dy \Bigg),\\
		\label{eqn_Laplace_BS_HD_U}
		\mathcal{L}_{\Phi_t^{\text{HD}}}(r,T) &= \exp\Bigg(-2 \pi p_{k,t}^{I} \int_{0}^{\Delta_k,t(r) + r} \lambda_{k,t}^{\text{BS,HD,U}}(y) y \left(1 - \mathcal{L}_h\left(\frac{P_t T r^{\alpha}}{P_u y^{\alpha}}\right)\right) dy \nonumber\\
		&\qquad\qquad\qquad -2 \pi \int_{\Delta_{k,t}(r) + r}^{\infty} \lambda_{k,t}^{\text{BS,HD,U}}(y) y \left(1 - \mathcal{L}_h\left(\frac{P_t T r^{\alpha}}{P_u y^{\alpha}}\right)\right) dy \Bigg),\\
		\label{eqn_Laplace_User_FD_U}
		\mathcal{L}_{\Phi_{u,t}^{\text{FD}}}(r, T) &= \exp\left(-2 \pi \int_{\max\{((\frac{P_t}{P_k})^{1 / \alpha} - 1)r - \delta_t, 0\}}^{\infty} \lambda_{k,t}^{\text{User,FD,U}}(y) y \left(1 - \mathcal{L}_h\left(\frac{T r^{\alpha}}{y^{\alpha}}\right)\right) dy \right),\\
		\label{eqn_Laplace_User_HD_U}
		\mathcal{L}_{\Phi_{u,t}^{\text{HD}}}(r, T) &= \exp\left(-2 \pi \int_{(\frac{P_k}{P_t})^{1/\alpha} \delta_t}^{\infty} \lambda_{k,t}^{\text{User,HD,U}}(y) y \left(1 - \mathcal{L}_h\left(\frac{T r^{\alpha}}{y^{\alpha}}\right)\right) dy \right),\\
		p_{k,t}^{I} &= 1 - \left(\frac{P_t^{1/ \alpha}}{P_t^{1/\alpha} + P_k^{1/\alpha}}\right)^2. 
		\end{align}
		\hrulefill
		\vspace*{4pt}
	\end{figure*}
\end{theorem}
\begin{IEEEproof}
	Please see Appendix \ref{appendix_SINR_U}.
\end{IEEEproof}

\begin{corollary}
\label{corollary_SINR}
The SINR performances of the users in conventional half-duplex HetNets outperform those in full-duplex HetNets, i.e.
\begin{align}
\label{eqn_corollary_1_D}
\mathcal{C}^{1\text{,HD}}(T) \vert_{\gamma_k = \infty} &\geq \left\{ \mathcal{C}^{i\text{,FD,D}}(T) \vert_{\gamma_k = 0} \right\}_{i = 1, 2}, \\
\label{eqn_corollary_1_U}
\mathcal{C}^{2\text{,HD}}(T) \vert_{\gamma_k = \infty} &\geq \left\{ \mathcal{C}^{i\text{,FD,U}}(T) \vert_{\gamma_k = 0} \right\}_{i = 1,2}.
\end{align}
\end{corollary}
\begin{IEEEproof}
	By plugging $\{\gamma_k = \infty\}_{k = 1, 2}$ and $\{\gamma_k = 0\}_{k = 1, 2}$ into Assumption \ref{assumption_intensity_downlink} and Theorem \ref{theorem_downlink_SINR_distribution}, we can easily have $\mathcal{L}_{\Phi_t^{\text{HD}}}|_{\gamma_k = \infty} \equiv \mathcal{L}_{\Phi_t^{\text{FD}}}|_{\gamma_k = 0}$, and 
	\begin{align}
	&\mathcal{C}_k^{1\text{,FD,D}}|_{\gamma_k = 0}(T) = \mathcal{C}_k^{2\text{,FD,D}}|_{\gamma_k = 0}(T) \nonumber\\
	&= \frac{2 \pi \lambda_k}{\mathcal{A}_k} \int\limits_{0}^{\infty} \exp\left(-\frac{T}{SSINR_{k,u}(r)} \right) \nonumber\\ 
	& \qquad \qquad \quad \times   \mathcal{F}(r) \prod\limits_{t = k, \bar{k}} \mathcal{L}_{\Phi_t^{\text{FD}}}(r,T) \mathcal{L}_{\Phi_{u,t}^{\text{FD}}}(r, T) dr,\\
	\label{eqn_half_duplex_only_D_SINR}
	&\mathcal{C}_k^{1\text{,HD}}|_{\gamma_k = \infty}(T) \nonumber\\
	&= \frac{2 \pi \lambda_k}{\mathcal{A}_k} \int\limits_{0}^{\infty} \exp\left(-\frac{T}{SNR_{k}(r)} \right) \mathcal{F}(r) \prod\limits_{t = k, \bar{k}}  \mathcal{L}_{\Phi_t^{\text{HD}}}(r, T) dr.
	\end{align}
	Since that $SSINR_{k,u}(r) \leq SNR_{k}(r)$ and $\mathcal{L}_{\Phi_{u,t}^{\text{FD}}}(r, T) \leq 1$, \eqref{eqn_corollary_1_D} is obtained.
	
	The proof of \eqref{eqn_corollary_1_U} is similar: 
	$\mathcal{L}_{\Phi_{u,t}^{\text{HD}}}|_{\gamma_k = \infty} \equiv \mathcal{L}_{\Phi_{u,t}^{\text{FD}}}|_{\gamma_k = 0}$,
	\begin{align}
	&\mathcal{C}_k^{1\text{,FD,U}}|_{\gamma_k = 0}(T) = \mathcal{C}_k^{2\text{,FD,U}}|_{\gamma_k = 0}(T) \nonumber\\
	&= \frac{2 \pi \lambda_k}{\mathcal{A}_k} \int\limits_0^{\infty} \exp\left(-\frac{T}{SSINR_{u,k}(r)}\right) \nonumber\\
	&\qquad \qquad \quad \times \mathcal{F}(r) \prod\limits_{t = k, \bar{k}} \mathcal{L}_{\Phi_t^{\text{FD}}}(r, T) \mathcal{L}_{\Phi_{u,t}^{\text{FD}}}(r, T) dr,\\
	\label{eqn_half_duplex_only_U_SINR}
	&\mathcal{C}_k^{2\text{,HD}}|_{\gamma_k = \infty}(T) \nonumber\\
	&= \frac{2 \pi \lambda_k}{\mathcal{A}_k} \int\limits_0^{\infty} \exp\left(-\frac{T}{SNR_u(r)}\right) \mathcal{F}(r) \prod\limits_{t = k, \bar{k}} \mathcal{L}_{\Phi_{u,t}^{\text{HD}}}(r, T) dr.
	\end{align}
\end{IEEEproof}

Corollary \ref{corollary_SINR} shows the impact of introducing full-duplex.
The RSI degrades the user SINR performances (e.g. $SSINR_{k,u}(r) \leq SNR_{k}(r)$).
However, even though the SIC is perfect, the inter-cell interference caused by full-duplex still degrades the users' SINRs (e.g. $\mathcal{L}_{\Phi_{u,t}^{\text{FD}}}(r, T) \leq 1$).

Generally, the transmit power of BSs is much greater than that of users.
Hence, in full-duplex HetNets ($\{\gamma_k\}_{k = 1,2} = 0$), the interferences from the full-duplex users could be ignorable compared to that from the BSs.
Then, we can have Corollary \ref{corollary_SINR_approximated}.
\begin{corollary}
	\label{corollary_SINR_approximated}
	When the SIC is perfect and the interferences from full-duplex users are neglected, the full-duplex downlink SINRs equal to the half-duplex downlink SINR.
\end{corollary}
\begin{IEEEproof}
	According to Theorem \ref{theorem_downlink_SINR_distribution} and Theorem \ref{theorem_uplink_SINR_distribution}, when $\{RSI_i\}_{i = 1,2,u} = 0$ and the interferences from full-duplex users are neglected, the approximated full-duplex downlink SINR distributions are given as 
	\begin{align}
	\label{eqn_full_duplex_only_D_SINR}
	&\widetilde{\mathcal{C}}_k^{1\text{,FD,D}}(T) \vert_{\gamma_k = 0} = \widetilde{\mathcal{C}}_k^{2\text{,FD,D}}(T) \vert_{\gamma_k = 0} \nonumber\\
	& = \frac{2\pi\lambda_k}{\mathcal{A}_k} \int\limits_{0}^{\infty} \exp\left(-\frac{T}{SNR_k(r)} \right) \mathcal{F}(r) \prod_{t=k,\bar{k}} \mathcal{L}_{\Phi_t^{\text{FD}}}(r,T) dr,
	\end{align}	
	One can see that  $\{\widetilde{\mathcal{C}}_k^{i\text{,FD,D}} \vert_{\gamma_k = 0} \}_{i = 1, 2} = \mathcal{C}_k^{1\text{,HD}}(T) \vert_{\gamma_k = \infty}$, i.e. the full-duplex downlink SINRs are equal to the half-duplex downlink SINR.	
\end{IEEEproof} 

Although Corollary \ref{corollary_SINR_approximated} for downlink SINR is easy to obtain, the corresponding corollary for uplink does not exist.
The approximated full-duplex uplink SINR distributions are given by
\begin{align}
\label{eqn_full_duplex_only_U_SINR}
&\widetilde{\mathcal{C}}_k^{1\text{,FD,U}}(T) \vert_{\gamma_k = 0} = \widetilde{\mathcal{C}}_k^{2\text{,FD,U}}(T) \vert_{\gamma_k = 0} \nonumber\\
& = \frac{2\pi\lambda_k}{\mathcal{A}_k} \int\limits_{0}^{\infty} \exp\left(-\frac{T}{SNR_u(r)} \right) \mathcal{F}(r) \prod_{t=k,\bar{k}} \mathcal{L}_{\Phi_t^{\text{FD}}}(r,T) dr,
\end{align} 
which are not equal to $\mathcal{C}_k^{2\text{,HD}}(T) \vert_{\gamma_k = \infty}$.


\subsection{Spectral Efficiency}
\label{subsec_SE_ER}
In this subsection, we derive the spectral efficiency of user in the proposed hybrid-duplex HetNet.
%
We denote $S_k^{i\text{,FD,D}}$ and $S_k^{i\text{,FD,U}}$ as the downlink and uplink full-duplex spectral efficiency on channel $i$ ($i \in \{1,2\}$) of the typical user in tier $k$, respectively. 
$S_k^{i\text{,HD}}$ is the half-duplex spectral efficiency on channel $i$.
\begin{align}
S_k^{i\text{,FD,D}}  &\triangleq \frac{1}{2} \mathbb{E}\left[\log_2 \left(1+\theta_k^{i\text{,FD,D}} \right) \right],\\
S_k^{i\text{,FD,U}} &\triangleq \frac{1}{2} \mathbb{E}\left[ \log_2 \left(1+\theta_k^{i\text{,FD,U}} \right) \right], \\
S_k^{i\text{,HD}} &\triangleq \frac{1}{2} \mathbb{E}\left[ \log_2 \left(1+\theta_k^{i\text{,HD}} \right)\right],
\end{align}
where `$\frac{1}{2}$' represents that the resources are divided into two orthogonal channel equally.
Hence, the spectral efficiency of the typical user is defined as
\begin{align}
& S \triangleq  \sum\limits_{i = 1,2} \sum\limits_{k = 1,2} \left( \mathcal{A}_k^{\text{FD}}  \left( S_k^{i\text{,FD,D}} + S_k^{i\text{,FD,U}} \right) +  \mathcal{A}_k^{\text{HD}} S_k^{i\text{,HD}} \right), 
\end{align}


\begin{theorem}
\label{theorem_rate_expectation}
The spectral efficiency is 
\begin{align}
\label{eqn_spectrum_efficiency}
& S = \frac{1}{2\ln 2} \Bigg(\sum\limits_{i = 1, 2} \sum\limits_{k=1,2} \mathcal{A}_k^{\text{FD}} \int\limits_{0}^{\infty}  \frac{\mathcal{C}_k^{i\text{,FD,D}}(T) + \mathcal{C}_k^{i\text{,FD,U}}(T)}{1+T}  dT \nonumber\\
&\qquad \qquad \quad + \sum\limits_{i = 1, 2} \sum\limits_{k=1,2} \mathcal{A}_k^{\text{HD}} \int\limits_{0}^{\infty} \frac{\mathcal{C}_k^{i\text{,HD}}(T)}{1+T} dT\Bigg).
\end{align}
\end{theorem}
\begin{IEEEproof}
Take $\mathbb{E}\left[S_k^{\text{FD,D}}\right]$ as example.
\begin{align}
S_k^{i\text{,FD,D}} &= \frac{1}{2} \mathbb{E}\left[ \log_2\left(1+\theta_k^{i\text{,FD,D}}\right)\right] \nonumber\\
&=\frac{1}{2 \ln 2} \int\limits_{0}^{\infty}\frac{1}{1+T} \mathcal{C}_k^{i\text{,FD,D}}(T)dT,
\end{align}
$S_k^{i\text{,FD,U}}$ and  $S_k^{i\text{,HD}}$ can be derived similarly.
Then the rate expectation \eqref{eqn_spectrum_efficiency} are obtained. 
\end{IEEEproof}

\begin{corollary}
\label{corollary_half_duplex_only_rate}
The spectral efficiency of the typical user in half-duplex two-tier HetNets is 
\begin{align}
& S = \frac{1}{2\ln 2} \sum\limits_{k=1,2} \mathcal{A}_k \int\limits_{0}^{\infty} \frac{\mathcal{C}_k^{1\text{,HD}}(T) \vert_{\gamma_k = \infty} + \mathcal{C}_k^{2\text{,HD}}(T) \vert_{\gamma_k = \infty}}{1+T} d \theta ,
\end{align}
where $\mathcal{C}_k^{i\text{,HD}}(\theta) \vert_{\gamma_k = \infty},i\in\{1,2\}$ are given as \eqref{eqn_half_duplex_only_D_SINR} and \eqref{eqn_half_duplex_only_U_SINR}, respectively. 
\end{corollary}

\begin{corollary}
\label{corollary_full_duplex_only_rate}
When the SIC is perfect and the interferences from full-duplex users are neglected, the average spectral efficiency of the typical user in full-duplex two-tier HetNets is
\begin{align}
& S = \frac{1}{\ln 2} \sum\limits_{k=1,2} \mathcal{A}_k \int\limits_{0}^{\infty}  \frac{\widetilde{\mathcal{C}}_k^{i\text{,FD,D}}(T) \vert_{\gamma_k = 0} + \widetilde{\mathcal{C}}_k^{i\text{,FD,U}}(T) \vert_{\gamma_k = 0}}{1+T}dT,
\end{align}
where $\widetilde{\mathcal{C}}_k^{i\text{,FD,D}}(T) \vert_{\gamma_k = 0}$ and $\widetilde{\mathcal{C}}_k^{i\text{,FD,U}}(T) \vert_{\gamma_k = 0}$ are given as \eqref{eqn_full_duplex_only_D_SINR} and \eqref{eqn_full_duplex_only_U_SINR}, respectively.
\end{corollary}

From Corollary \ref{corollary_half_duplex_only_rate} and Corollary \ref{corollary_full_duplex_only_rate}, one can see that the amount of resource doubles for both downlink and uplink transmit in full-duplex HetNets.
Especially, as $\{\widetilde{\mathcal{C}}_k^{i\text{,FD,D}}(T) \vert_{\gamma_k = 0}\} = \mathcal{C}_k^{1\text{,HD}}(T) \vert_{\gamma_k = \infty}$, the downlink spectral efficiency is doubled when the SIC is perfect and the interferences from the full-duplex users are neglected.
However, the uplink performance is much more complicated to analyze.
More insights about the spectral efficiency are shown in Sec. \ref{sec_numerical_results}. 

\section{Received-Power Thresholds Optimization}
\label{sec_system_optimization}
In the previous section, we derive the analytical performance for given received power thresholds $\{\gamma_k\}_{k = 1,2}$.
In this section, we try to optimize the received power thresholds for sum rate maximization. 

\subsection{Problem Formulation}
There are $M$ cells in the HetNet.
In cell $m$, there are one BS and $N_m$ users.
Each BS schedules users independently with round-robin policy. 
$P_{m, u}$ is denoted as the average downlink received power of user $u$ in cell $m$.
The average uplink received power of user $u$ in cell $m$ is denoted as $P_{u, m}$.
\footnote{Here, we consider that the average received powers are determined by transmit power and path loss. 
Hence, the average received powers are the same on both channels.}
Without loss of generality, the users belonging to a BS are sorted in decrease of the average downlink received power, i.e. 
\begin{align}
P_{m,u} \geq P_{m,v}, 1 \leq u<v\leq N_m.
\end{align}

Denote $I_{n,m,0}$ and $I_{n,m,u}$ as the interference from the BS in cell $n$ to the BS and user $u$ in cell $m$, respectively.
For simplicity, we assume that the interferences from any user in cell $n$ to the BS in cell $m$ are the same, which is denoted as $I_{n,m,0}' $.
Similarly, the interference from any user in cell $n$ to user $u$ in cell $m$ is denoted as $I_{n,m,u}'$.
\footnote{This assumption is to eliminate the impact caused by the scheduling order of users.
It helps to provide stable received-power thresholds. 
The scheduling problem considering the exact interference in hybrid-duplex HetNets is an interesting work, but it is beyond the scope of this paper.
To obtain $\{I_{n, m, 0}'\}$ and $\{I_{n, m, u}'\}$, in Sec. \ref{sec_numerical_results}, we assume that $I_{n, m, 0}' = \epsilon_m I_{n, m. 0}$ and $I_{n, m, u}' = \epsilon_m I_{n, m, u}$, where $\epsilon_m$ is the user transmit power to the BS transmit power ratio in cell $m$.
It is worth noting that, in Sec. \ref{sec_numerical_results}, the assumption is used in the optimization process \emph{only}.
The simulations of system performance do \emph{not} adopt this assumption.
The simulation results verify that our algorithm still provides a great improvement with this assumption.}

Let $\Delta_m \in [0, N_m]$ be the number of full-duplex users in BS $m$.
If $u \leq \Delta_m$, user $u$ is scheduled as a full-duplex user, otherwise as a half-duplex user.
Then the probability that BS $m$ is in full-duplex mode is $\frac{\Delta_m}{N_m}$.
Hence, the rate of user $u$ can be expressed as \eqref{eqn_opti_rate_FD_D}, \eqref{eqn_opti_rate_FD_U} and \eqref{eqn_opti_rate_HD} on the top of page $8$, respectively.
\begin{figure*}[t]
	\normalsize
	\begin{align}
	\label{eqn_opti_rate_FD_D}
	&R_{m,u}^{\text{FD,D}} =  \log_2 \left(\frac{P_{m,u}}{RSI_u + \sigma^2 + \sum\limits_{n \in M \setminus m} \left(I_{n,m,u} + \frac{\Delta_n}{N_n}I_{n,m,u}'\right)} \right) + \log_2 \left(\frac{P_{m,u}}{RSI_u + \sigma^2 + \sum\limits_{n \in M \setminus m} \left(\frac{\Delta_n}{N_n} I_{n,m,u} + I_{n,m,u}'\right)} \right),\\
	\label{eqn_opti_rate_FD_U}
	&R_{m,u}^{\text{FD,U}} =  \log_2 \left(\frac{P_{u,m}}{RSI_m + \sigma^2 + \sum\limits_{n \in M \setminus m} \left(I_{n,m,0} + \frac{\Delta_n}{N_n}I_{n,m,0}'\right)} \right) + \log_2 \left(\frac{P_{u,m}}{RSI_m + \sigma^2 + \sum\limits_{n \in M \setminus m} \left(\frac{\Delta_n}{N_n} I_{n,m,0} + I_{n,m,0}'\right)} \right),\\
	\label{eqn_opti_rate_HD}
	&R_{m,u}^{\text{HD}} =  \log_2 \left(\frac{P_{m,u}}{\sigma^2 + \sum\limits_{n \in M \setminus m} \left(I_{n,m,u} + \frac{\Delta_n}{N_n}I_{n,m,u}'\right)} \right) + \log_2 \left(\frac{P_{u,m}}{\sigma^2 + \sum\limits_{n \in M \setminus m} \left(\frac{\Delta_n}{N_n} I_{n,m,0} + I_{n,m,0}'\right)} \right).
	\end{align}
	\hrulefill
	\vspace*{4pt}
\end{figure*}

Let $\omega_m = \frac{W}{N_m}$.
To maximize the sum rate of the users, we have the following optimization problem:
\begin{align}
\label{eqn_opt_obj}
&\max_{\{ \Delta_m \}} U = \sum\limits_{m \in M} \omega_m \Big( R_{m}^{\text{FD}} + R_{m}^{\text{HD}} \Big), \\
&\text{s.t.} \qquad \forall m \in M, \Delta_m \in [0, \dots, N_m],\\
&\qquad \quad R_{m}^{\text{FD}} = 
\begin{cases}
0, \text{ if } \Delta_m = 0,\\
\sum\limits_{u = 1}^{\Delta_m} \left(R_{m,u}^{\text{FD,D}} + R_{m,u}^{\text{FD,U}}\right), \text{otherwise},
\end{cases}\\
&\qquad \quad R_{m}^{\text{HD}} = 
\begin{cases}
0, \text{ if } \Delta_m = N_m,\\
\sum\limits_{u = \Delta_m + 1}^{N_m} R_{m,u}^{\text{HD}}, \text{otherwise}. 
\end{cases}
\end{align}

\subsection{Suboptimal Algorithm}
\begin{algorithm}
	\centering
	\caption{Duplex Mode Selection Algorithm}
	\label{alg_duplex_threshold}
	\begin{algorithmic}[1]
		\STATE Input: \\
		system parameters: $W$, $\sigma^2$; $\{N_m, RSI_m, RSI_u\}$, \\$\forall u \in [1,N_m]$, $\forall m \in M$; \\
		measurements: $\{I_{n,m,u}, I_{n,m,0}, I_{n,m,u}', I_{n,m,0}'\}$,  \\$\{P_{m,u}, P_{u,m}\}$, $\forall u \in [1,N_m], \forall n, m \in M, n \neq m$.\\
		\STATE Initialize $\bm{\Delta}^* = \bm{0}$; Compute $U^*$ using \eqref{eqn_opt_obj}; $U^+ = U^*$, $U^- = U^*$;
		\LOOP
		\FOR {$m = 1$ to $M$}	    
		\IF {$\Delta_m^* + 1 \leq N_m$}
		\STATE $\bm{\Delta}' = \bm{\Delta}^*$; $\Delta_m' = \Delta_m' + 1$; Compute $U'$ by \eqref{eqn_opt_obj};
		\IF{$U' > U^+$}
		\STATE $\bm{\Delta}^+ = \bm{\Delta}'$, $U^+ = U'$
		\ENDIF
		\ENDIF
		\IF {$\Delta_m^* - 1 \geq 0$}
		\STATE $\bm{\Delta}' = \bm{\Delta}^*$; $\Delta_m' = \Delta_m' - 1$; Compute $U'$ by \eqref{eqn_opt_obj};
		\IF{$U' > U^-$}
		\STATE $\bm{\Delta}^- = \bm{\Delta}'$; $U^- = U'$
		\ENDIF
		\ENDIF
		\ENDFOR
		\IF{$U^+ \geq U^- \text{ and } U^+ > U^*$}
		\STATE $U^* = U^+$; $\bm{\Delta}^* = \bm{\Delta}^+$;
		\ELSE
		\IF{$U^- \geq U^+ \text{ and } U^- > U^*$}
		\STATE $U^* = U^-$; $\bm{\Delta}^* = \bm{\Delta}^-$;
		\ELSE
		\RETURN $\bm{\Delta}^*$;
		\ENDIF
		\ENDIF
		\ENDLOOP
	\end{algorithmic}
\end{algorithm}

In this subsection, we propose a centralized greedy algorithm to achieve a suboptimal solution, which is presented in Algorithm \ref{alg_duplex_threshold}.
First, all nodes measure the received signal and interference power and feed them back to the algorithm (Line 1). 
The centralized algorithm initializes the vector $\bm{\Delta} \triangleq \{\Delta_m\}$ and computes the sum rate (Line 2).
Then, we find the BS with maximum sum rate if one more/less full-duplex user served by it (Line 4 to 17). 
Finally, the algorithm updates the vector $\bm{\Delta}^*$ with the largest sum rate if the system performance improves (Line 18 to 26).
The algorithm ends when there is no improvement. 

By using Algorithm \ref{alg_duplex_threshold}, we have the suboptimal vector $\bm{\Delta}^*$.
Then we could compute the received power thresholds for each BS as 
\begin{align}
\gamma_m = 
\begin{cases}
0, \text{if } \Delta_m^* = 0,\\
\frac{P_{m,\Delta_m} + P_{m,\Delta_m+1}}{2}, \text{if } 0 < \Delta_m^* < N_m,\\
\infty, \text{if } \Delta_m^* = N_m.
\end{cases}
\end{align}

\begin{figure}[t]
	\centering
	\includegraphics[width=85mm]{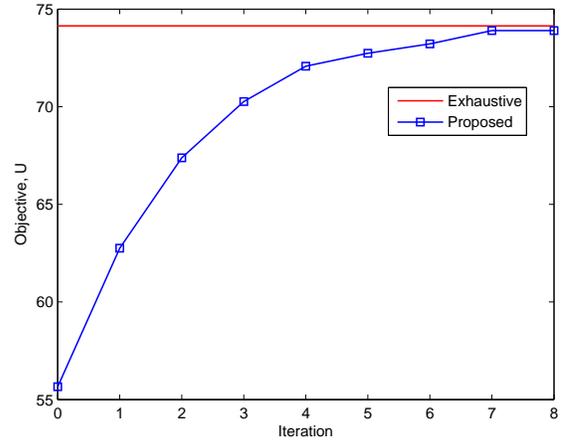}
	\vspace{-0.1cm}
	\caption{The iteration and performance of Algorithm \ref{alg_duplex_threshold}.}
	\label{fig_iteration}
	\vspace{-0.3cm}
\end{figure}

As the optimization problem is a nonlinear integer programming, it is difficult to obtain the optimal solution in polynomial time. 
Thus we compare the proposed suboptimal algorithm with the optimal solution by exhaustive search with small number of cells and users in Fig. \ref{fig_iteration}, where we set $13$ cells and the numbers of users in each cell are $[6,6,8,3,8,4,4,1,1,2,1,1,1]$ respectively.
Our algorithm computes $2M$ times of the objective in each iteration, while the total number of computation in exhaustive search is $\prod\limits_{m \in M}(N_m + 1)$.
In Fig. \ref{fig_iteration} , we observe that Algorithm 1 performs closely to the exhaustive search.
Since the proposed algorithm increases the objective in each iteration and thus it converges fast.


\section{Numerical Results}
\label{sec_numerical_results}
In this section, we verify the theoretical analysis and demonstrate the effectiveness of our proposed optimization algorithm by using numerical results.
\subsection{Validation and Insights of the Theoretical Analysis}
In this subsection, we verify our theoretical analysis. 
We set the densities of two tiers as $\lambda_1 = 1 \text{ BS}/\text{km}^2$ and $\lambda_2 = 10 \text{ BS}/\text{km}^2$.
The user density is $\lambda_u = 50 \text{ users}/\text{km}^2$.
The path loss exponent $\alpha$ is assumed to be $3.5$.
The transmit powers are $P_1 = 46$ dBm, $P_2 = 30$ dBm and $P_u = 23$ dBm, respectively.
We assume a linear RSI performance, i.e., $\{RSI_t = \beta P_t\}_{t=1,2,u}$ \cite{ramirez_optimal_2013}, and the RSI ratio $\beta = -70$ dB.
The bandwidth $W$ is assumed to be $20$ MHz ($10$ MHz for downlink and uplink channel, respectively) and the thermal noise power density is $-174$ dBm$/$Hz.
%
%
The received power thresholds are set to $\gamma_1 = -71$ dB and $\gamma_2 = -76$ dB, respectively.
All the results in this subsection are obtained under the parameter settings mentioned above, except where otherwise noted. 

\begin{figure}[t]
	\centering
	\includegraphics[width=85mm]{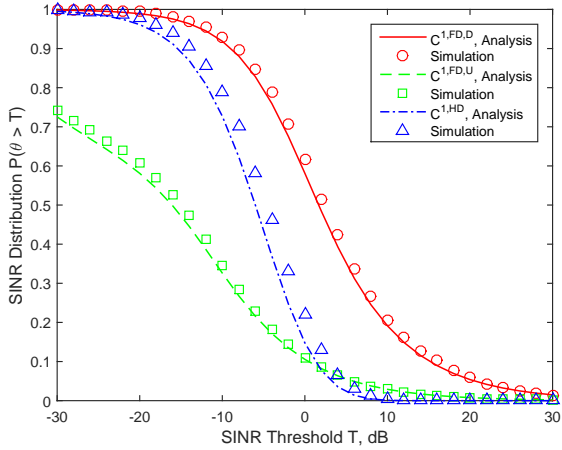}
	\vspace{-0.1cm}
	\caption{The SINR distributions on channel $1$, theoretical analysis VS simulation.}
	\label{fig_SINR_Chn_1}
	\vspace{-0.3cm}
\end{figure}

\begin{figure}[t]
	\centering
	\includegraphics[width=85mm]{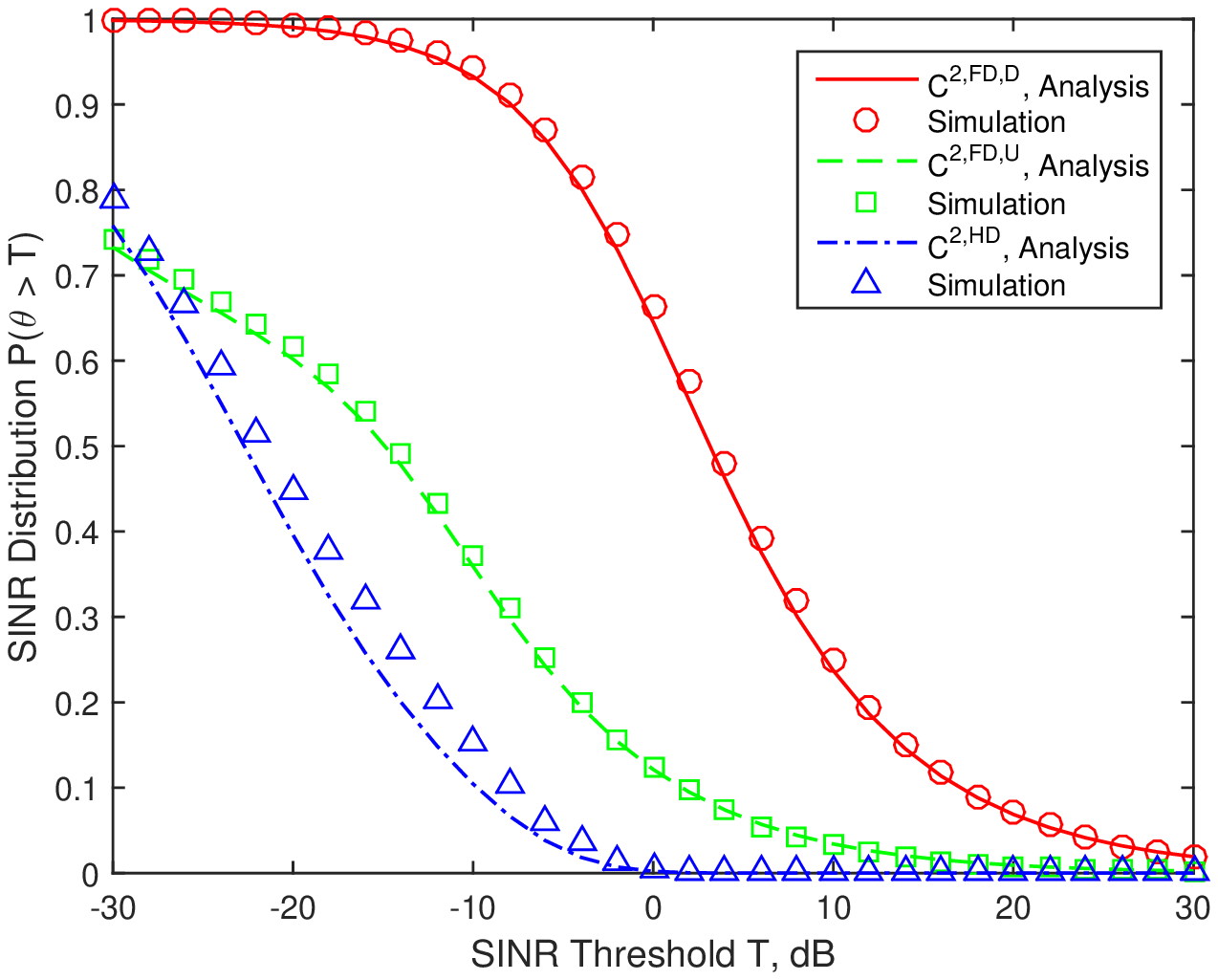}
	\vspace{-0.1cm}
	\caption{The SINR distributions on channel $2$, theoretical analysis VS simulation.}
	\label{fig_SINR_Chn_2}
	\vspace{-0.3cm}
\end{figure}

In Fig. \ref{fig_SINR_Chn_1} and Fig. \ref{fig_SINR_Chn_2}, the SINR distributions of Theorem \ref{theorem_downlink_SINR_distribution}, 
\ref{theorem_uplink_SINR_distribution} and that obtained through Monte Carlo simulation are shown for channel $1$ and channel $2$, respectively.
One can see that our analysis results show a good agreement with the simulations in all the cases.
It also verifies the rationality of the assumptions and approximations which we use in the theoretical derivations.
The SINR performance of the half-duplex users is inferior to the full-duplex users (i.e. $\mathcal{C}^{1\text{,FD,D}}$ and $\mathcal{C}^{1\text{,HD}}$ in Fig. \ref{fig_SINR_Chn_1}, $\mathcal{C}^{2\text{,FD,U}}$ and $\mathcal{C}^{2\text{,HD}}$ in Fig. \ref{fig_SINR_Chn_2}), although they do not suffer RSI.
It is because that the half-duplex users have lower received power (less than $\gamma_k$) than the full-duplex users.

\begin{figure}[t]
	\centering
	\includegraphics[width=85mm]{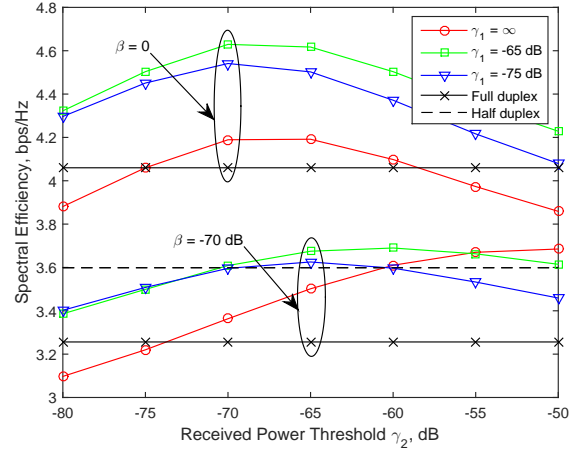}
	\vspace{-0.1cm}
	\caption{The spectral efficiencies with varied received power thresholds; perfect and imperfect SIC.}
	\label{fig_SE_Total}
	\vspace{-0.3cm}
\end{figure}
Fig. \ref{fig_SE_Total} shows the spectral efficiencies with varied received power thresholds.
RSI has a great impact on the spectral efficiencies.
With perfect SIC (i.e. $\beta = 0$), the best spectral efficiency is much better than that in the half-duplex mode.
But the spectral efficiency performance degrades greatly when the SIC capacity decreases to $-70$ dB, and the full-duplex performance is even worse than the half-duplex case.
Furthermore, even with perfect SIC, the improvement of spectral efficiency is still far from doubling.
It is because that using full-duplex increases inter-cell interference.
We can observe that the best spectral efficiency performance is obtained in the proposed scheme.
There is a tradeoff between amount of available resources and inter-cell interference.
Our proposed received-power-based hybrid-duplex policy can provide a better balance than the full/half-duplex cases, and it provides better performance. 

\begin{figure}[t]
	\centering
	\includegraphics[width=85mm]{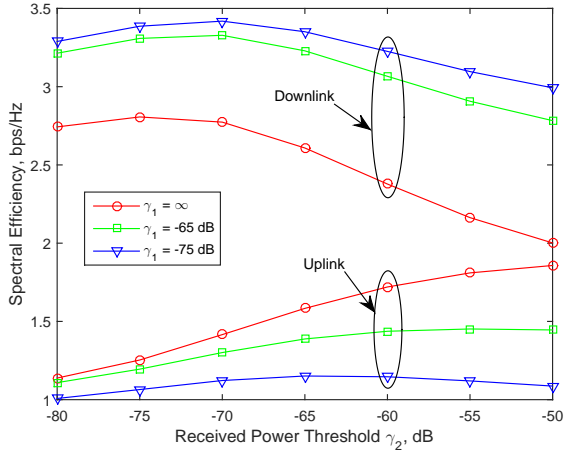}
	\vspace{-0.1cm}
	\caption{The downlink/uplink spectral efficiencies with varied received power thresholds; perfect SIC.}
	\label{fig_SE_D_U}
	\vspace{-0.3cm}
\end{figure}
\begin{figure}[t]
	\centering
	\includegraphics[width=85mm]{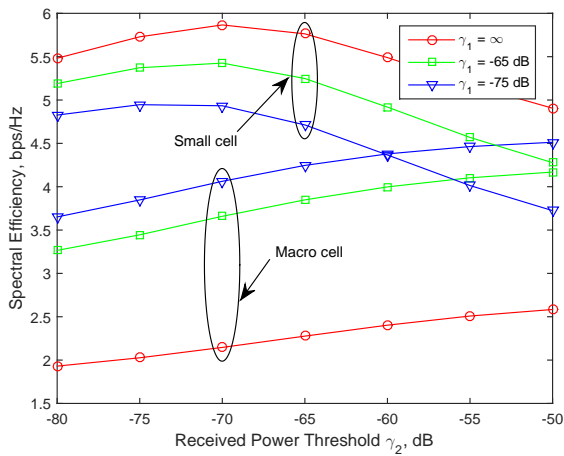}
	\vspace{-0.1cm}
	\caption{The spectral efficiencies in two tiers with varied received power thresholds; perfect SIC.}
	\label{fig_SE_M_P}
	\vspace{-0.3cm}
\end{figure}
The spectral efficiencies of downlink and uplink with perfect SIC are shown in Fig. \ref{fig_SE_D_U}.
The downlink and uplink spectral efficiencies are defined as
\begin{align}
& S^{\text{D}} \triangleq  \sum\limits_{k = 1,2} \Big( \mathcal{A}_k^{\text{FD}} \sum\limits_{i = 1,2} S_k^{i\text{,FD,D}} +  \mathcal{A}_k^{\text{HD}} S_k^{1\text{,HD}} \Big),\\
& S^{\text{U}} \triangleq  \sum\limits_{k = 1,2} \Big( \mathcal{A}_k^{\text{FD}} \sum\limits_{i = 1,2} S_k^{i\text{,FD,U}} +  \mathcal{A}_k^{\text{HD}} S_k^{2\text{,HD}} \Big).
\end{align}
The numerical results verify the discussion about Corollary \ref{corollary_half_duplex_only_rate} and Corollary \ref{corollary_full_duplex_only_rate} in Sec. \ref{subsec_SE_ER}.
Full duplex benefits the downlink spectral efficiency performance.
As the threshold $\gamma_1$ decreases, the downlink spectral efficiency improves.
On the other hand, as more macro cells transmit on the uplink channel (i.e. $\gamma_1$ decreases), the interference level increases and the uplink performance of users degrades.
In a way, employing full-duplex improves the downlink spectral efficiency by sacrificing the uplink performance.

In Fig. \ref{fig_SE_M_P}, we define the spectral efficiencies of the macro users and the small cell users as
\begin{align}
& S_{k} \triangleq \frac{1}{\mathcal{A}_k} \sum\limits_{i = 1,2} \Big( \mathcal{A}_k^{\text{FD}} \left( S_k^{i\text{,FD,D}} + S_k^{i\text{,FD,U}} \right) +  \mathcal{A}_k^{\text{HD}} S_k^{i\text{,HD}} \Big).
\end{align}
As $\gamma_2$ decreases from $-50 \text{ dB}$ to $-60 \text{ dB}$, the spectral efficiency of the users in small cells grows greatly while the spectral efficiency of the macro users has a slight impact.
Compared with the macro cells,  the transmit power of small cells is much lower.
This character helps to restrain the inter-cell interference caused by the use of full-duplex in small cells.
This justifies the intuition that low-power small cells are more suitable candidate to deploy full-duplex.


\subsection{Performance of the Optimization of Received Power Thresholds}
In this subsection, we show the performance of the proposed greedy algorithm.
We set a $1 \text{ km}^2$ area and the path loss exponent is $\alpha = 3.5$.
The two-tier BSs and the users are drawn from independent PPPs, respectively.
The macro BS density is $\lambda_1 = 1 \text{ BS}/\text{km}^2$ and the small cell density varies.
The user PPP has a density of $\lambda_u = 150 \text{ users} / \text{km}^2$.
The system bandwidth and the transmit power of the network elements are assumed to be the same as in previous subsection. 
We perform Monte Carlo simulation over $200$ snapshots of different spatial topologies.
Each snapshot consists of $1,000$ time slots.

\begin{figure}[t]
	\centering
	\includegraphics[width=85mm]{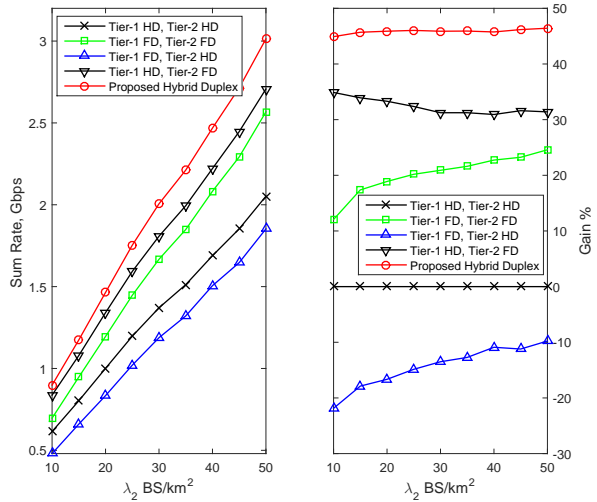}
	\vspace{-0.1cm}
	\caption{Effectiveness of optimization with varied small cell density. The left figure shows the sum rate of users, and the right figure shows the gain over the conventional half-duplex HeNets.}
	\label{fig_opt_performance}
	\vspace{-0.3cm}
\end{figure}
According to \cite{lee_hybrid_2015}, operating all BSs in full-duplex or half-duplex mode achieves higher throughput compared to the mixture of two mode BSs in each tier.
Hence, we consider four duplex mode sets as benchmarks: 1) both tiers using half-duplex mode, 2) both tiers using full-duplex mode, 3) macro cells in full-duplex mode while small cells in half-duplex mode and 4) macro cells in half-duplex mode while small cells in full-duplex mode. 
Due to computational complexity, the exhaustive search results are omitted here.
The sum rate obtained from different duplex mode sets are shown on the left of Fig. \ref{fig_opt_performance}.
We also show the gain of sum rate compared to the conventional half-duplex HetNets on the right.
As observed in Fig. \ref{fig_opt_performance}, our proposed hybrid-duplex mode outperforms all the benchmarks.
As the density of small cells grows, the system performance of all the duplex mode sets improve.
Higher density of BSs benefits the system performance in two aspects.
First, the average load of BS decreases as the density grows, and the weights $\{\omega_m\}$ increase.
Second, the average transmitting distance decreases as the BS density grows.
However, a higher density of BSs causes stronger inter-cell interference, especially when small cells use full-duplex.
It is the reason why the gain decreases in the case that macro cells use half-duplex and small cells use full-duplex in the right figure.
On the other hand, the proposed algorithm shows a good balance between the duplex mode selection and interference suppression.
The system gain stabilizes at about $45 \%$ for all densities.
It verifies that our algorithm can fulfill the potential of the proposed hybrid-duplex HetNets.

\section{Conclusion}
\label{sec_conclusion}
This paper proposes a novel received-power-based hybrid-duplex HetNet scheme.
We characterize the SINR distribution and spectral efficiency by using stochastic geometry.
Especially, inhomogeneous PPP model is used for both downlink and uplink analysis.
A greedy algorithm for optimizing received power thresholds is also proposed. 
Extensive performance evaluations are conducted for both the analytical model and the optimization algorithm.
The results show that the proposed hybrid-duplex HetNet scheme can improve the system performance significantly. 


\appendices

\section{}
\label{appendix_policy}
We consider a typical user in a two-tier HetNet.
Define the following function:
\begin{align}
\label{eqn_appendix_a}
f\left(\xi\right) &=  \overbrace{\frac{1}{2} \log\left(1 + \frac{\xi}{N_{\text{D}}}\right) + \frac{1}{2} \log\left(1 + \frac{\epsilon \xi}{N_{\text{U}}}\right)}^{\text{Half-duplex}} \nonumber \\
& \underbrace{- \log\left(1 + \frac{\xi}{RSI_{\text{D}} + N_{\text{D}}}\right) - \log\left(1 + \frac{\epsilon \xi}{RSI_{\text{U}} + N_{\text{U}}}\right)}_{\text{Full-duplex}},
\end{align} 
where $\xi$ is the average downlink received power.
$\epsilon$ is the user transmit power to the BS transmit power ratio.
Note that the path losses are the same on both downlink and uplink channel.
$N_{\text{D}}$ and $N_{\text{U}}$ are denoted as the sum of interference and noise on the downlink and uplink channel.
$RSI_{\text{D}}$ and $RSI_{\text{U}}$ are the residual self-interference of the user and its serving BS, respectively.

Eqn. \eqref{eqn_appendix_a} is the difference of data rate between the half-duplex mode and the full-duplex mode.
When $f\left(\xi\right) \leq 0$, it means that the user has better performance in full-duplex mode.
Otherwise, half-duplex mode is the better choice when $f\left(\xi\right) > 0$. 

When $RSI_{\text{D}} \geq N_{\text{D}}$ and $RSI_{\text{U}} \geq N_{\text{U}}$, it is easy to have the following properties about $f\left(\xi\right)$ :
\begin{enumerate}[1)]
	\item $f\left(0\right) = 0$;
	\item when $\xi$ is sufficient small (but greater than $0$), the gradient $\frac{d f(\xi)}{d \xi} > 0$;
	\item when $\xi > \max\{RSI_{\text{D}} - N_{\text{D}}, \frac{1}{\epsilon}\left(RSI_{\text{U}} - N_{\text{U}}\right)\}$, the gradient $\frac{d f(\xi)}{d \xi} < 0$;
	\item $\lim\limits_{\xi \to \infty} f\left(\xi\right) < 0$.
\end{enumerate}

Hence, when SIC is not perfect and RSIs are relatively strong, a user suffering low received signal power ($\xi$ is sufficient small) should be scheduled as a half-duplex user ($f\left(\xi\right) > 0$).
Otherwise, the user should be scheduled as a full-duplex user ($f\left(\xi\right) < 0$) when the received signal power is sufficient large.
This is the reason why we consider a hybrid-duplex scheme based on average downlink received signal power.

\section{}
\label{appendix_association}
According to the user association policy \eqref{eqn_association_policy}, the probability that the typical user is a full-duplex user in tier $k$ is
\begin{align}
&\mathcal{A}_k^{\text{FD}} = \mathbb{P} \left(D_k \leq D_{\bar{k}} \left(\frac{P_k}{P_{\bar{k}}} \right)^{1/\alpha} \cap D_k \leq \delta_k \right) \nonumber\\
&= 1 - \int\limits_{0}^{\mu_k(\delta_k)} \mathbb{P} \left( D_k > r \left( \frac{P_k}{P_{\bar{k}}} \right)^{1/\alpha} \right) f_{D_{\bar{k}}}(r) dr \nonumber\\
&\qquad \qquad - \int\limits_{\mu_{k}(\delta_k)}^{\infty} \mathbb{P} \left( D_k > \delta_k \right) f_{D_{\bar{k}}}(r) dr, 
\end{align}
where $\delta_k \triangleq \left(\frac{P_k}{\gamma_k}\right)^{1/\alpha}$ and $\mu_k(x) \triangleq x \left(\frac{P_{\bar{k}}}{P_k} \right)^{1/\alpha}$.

Using the property of PPP \cite{baccelli_stochastic_2009}, we have the full-duplex association probability \eqref{eqn_FD_association_probability}.
Following a similar process, the half-duplex association probability \eqref{eqn_HD_association_probability} is derived. 
And \eqref{eqn_tier_k_association_probability} follows by the definition.

\section{}
\label{appendix_intensity}
Take the point process of interfering full-duplex BSs $\Phi_t^{\text{FD}}$ as example.
\begin{figure}[t]
	\centering
	\includegraphics[width=85mm]{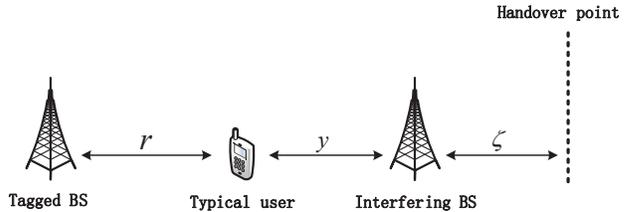}
	\vspace{-0.1cm}
	\caption{An instance of the distances between nodes.}
	\label{fig_intensity}
	\vspace{-0.3cm}
\end{figure}
As shown in Fig. \ref{fig_intensity}, $r$ denotes as the distance between the typical user and the tagged BS.
$y$ is the distance from the interfering BS to the typical user.
Let $\zeta$ be the distance between the interfering BS and the handover point.
According to the association policy, we have
\begin{equation}
\zeta = \frac{r + y}{\max\{(\frac{P_k}{P_t})^{1/\alpha} - 1, 0\}}.
\end{equation}
We assume that the coverage area of the interfering BS is a circular region around it with radius $\zeta$.
Then, the probability that there is user located in the coverage area, i.e. the probability that the interfering BS is active, can be expressed as $1 - \exp\left(-\lambda_u \pi \zeta^2\right)$.
Thus the intensity measure function \eqref{eqn_intensity_BS_FD_D} is obtained.
The process of deriving \eqref{eqn_intensity_BS_HD_D} is similar except that the interfering BS is scheduling a half-duplex user, which means that $\zeta$ should be greater than $\delta_t$.
Intensity functions \eqref{eqn_intensity_User_FD_D} and \eqref{eqn_intensity_User_HD_D} follow the discussion in \cite{singh_joint_2015}. 

It is worth mentioning that all the intensity measure functions in this paper are the upper bound on accurate inhomogeneous intensities.
Hence, the theoretical SINR distributions are lower-bounds, which is verified in Section \ref{sec_numerical_results}.

\section{}
\label{appendix_SINR_D}
Take the downlink full-duplex SINR distribution $\mathcal{C}_k^{1\text{,FD,D}}$ as example.
\begin{align}
\label{eqn_SINR_FD_D_decompose}
\mathcal{C}_k^{1\text{,FD,D}}(T) = \int\limits_{0}^{\infty} \mathbb{P}(\theta_k^{1\text{,FD,D}}(r) > T | D_k^{\text{FD}} = r) f_{D_k^{\text{FD}}}(r) dr.
\end{align}
According to the association policy \eqref{eqn_association_policy}, the cumulative density function (CDF) of the distance $D_k^{\text{FD}}$ is:
\begin{equation}
\mathbb{P}(D_k^{\text{FD}} \leq x) = 
\begin{cases}
1,x>\delta_k,\\
\mathbb{P}(D_k \leq x | u \in \mathcal{U}_k^{\text{FD}}), x \leq \delta_k,
\end{cases}
\end{equation}
where
\begin{align*}
&\mathbb{P}(D_k \leq x | u \in \mathcal{U}_k^{\text{FD}}) = \frac{\mathbb{P}(D_k \leq x, u \in \mathcal{U}_k^{\text{FD}})}{\mathbb{P}(u \in \mathcal{U}_k^{\text{FD}})}\nonumber\\
&=\frac{1}{\mathcal{A}_k^{\text{FD}}} \bigg[1-\exp\left(-\pi \lambda_k x^2 -\pi \lambda_{\bar{k}}(\mu_k(x))^2 \right) \nonumber\\
&\qquad -2\pi \lambda_{\bar{k}} \int\limits_{0}^{\mu_k(x)} r \exp \left(- \pi \lambda_k \Big(\frac{P_k}{P_{\bar{k}}} \Big)^{\frac{2}{\alpha}} r^2 -\pi \lambda_{\bar{k}} r^2\right)dr
\bigg].
\end{align*}
Thus the probability density function (pdf) is:
\begin{align}
\label{eqn_R_k_FD_pdf}
&f_{D_k^{\text{FD}}}(x) = 
\begin{cases}
0,x>\delta_k,\\
\frac{2\pi \lambda_k}{\mathcal{A}_k^{\text{FD}}}\left[ x \exp \left(-\pi \lambda_k x^2 -\pi \lambda_{\bar{k}}(\mu_k(x))^2\right)\right],x \leq \delta_k. 
\end{cases}
\end{align}
The conditional SINR distribution on channel $1$ of a full-duplex user at distance $r$ from its serving BS is
\begin{align}
\label{eqn_conditional_theta_k_FD_D}
&\mathbb{P}\left(\theta_k^{1\text{,FD,D}} > T | D_k^{\text{FD}} = r \right) \nonumber\\
&=\exp \left(-\frac{T}{SSINR_{k,u}(r)}\right) \prod\limits_{t=k,\bar{k}} \mathbb{E}_{I_{t,k}^{1\text{,D}}} \left[\exp\left(-\frac{T r^{\alpha}}{P_k} I_{t,k}^{1\text{,D}} \right) \right],
\end{align}

Using the interference expression \eqref{eqn_channel_1_downlink_interference}, we have
\begin{align}
\label{eqn_interference_D_expectation}
&\mathbb{E}_{I_{t,k}^{1\text{,D}}} \left[\exp\left(- \frac{T r^{\alpha}}{P_k} I_{t,k}^{1\text{,D}} \right) \right] \nonumber\\
&\qquad \qquad \overset{(a)}{=} \mathcal{L}_{\Phi_t^{\text{FD}}}(r,T) \mathcal{L}_{\Phi_{u,t}^{\text{FD}}}(r, T) \mathcal{L}_{\Phi_t^{\text{HD}}}(r, T),
\end{align} 
where $(a)$ follows that $\Phi_t^{\text{FD}}$, $\Phi_{u,t}^{\text{FD}}$ and $\Phi_t^{\text{HD}}$ are assumed to be mutually independent, $h_{b_i, u_0}$ and $h_{u_{b_i}, u_0}$ are i.i.d. ($h_{i,j} \sim \exp(1)$).
Using the probability generating functional (PGFL) \cite{stoyan_stochastic_2013} of PPP and the Laplace transform $\mathcal{L}_h(s) = \frac{1}{1+s}$ of $h \sim \exp(1)$, we have \eqref{eqn_Laplace_BS_FD_D} and \eqref{eqn_Laplace_BS_HD_D}.
$\Delta_{k,t}(r) = \left(\frac{P_t}{P_k} \right)^{1/\alpha} r$ is the lower-bound of the distance to the interfering BS in tier $t$ when the typical user is in tier $k$.

The derivation of \eqref{eqn_Laplace_User_FD_D} is similar except the interesting lower-bound $\max\{r(\frac{P_t}{P_k})^{1 / \alpha} - \delta_t, 0\}$.
Let $\zeta$ be the distance between the interfering full-duplex user and its serving BS.
Remind that $r$ is the distance between the typical user and the tagged BS, and $y$ denotes as the distance from the interfering user to the typical one.
There are two critical conditions: (i) the interfering user is full-duplex; (ii) the typical user is served by the tagged BS.
Hence we have
\begin{align}
\begin{cases}
\zeta \leq \delta_t\\
P_k r^{-\alpha} \geq P_t(y+\zeta)^{-\alpha} 
\end{cases}
\Rightarrow y \geq r(\frac{P_t}{P_k})^{1 / \alpha} - \delta_t,
\end{align}
and then \eqref{eqn_Laplace_User_FD_D} is obtained.

The lower-bound $\max\{\delta_t(\frac{P_k}{P_t})^{1 / \alpha} - r, 0\}$ in \eqref{eqn_Laplace_User_HD_D} is obtained considering two different conditions: (i) the interfering user is half-duplex; (ii) the interfering user is not served by the tagged BS.
Hence we have
\begin{align}
\begin{cases}
\zeta \geq \delta_t\\
P_k (r + y)^{-\alpha} \leq P_t \zeta^{-\alpha}
\end{cases}
\Rightarrow y \geq \delta_t (\frac{P_k}{P_t})^{1 / \alpha} - r.
\end{align}

The proofs of \eqref{eqn_channel_1_SINR_HD} and \eqref{eqn_channel_2_SINR_FD_D} are omitted, which can be derived following a similar process.

\section{}
\label{appendix_SINR_U}
The proof of the uplink SINR distributions is roughly the same as that of downlink.
Similar to Appendix \ref{appendix_SINR_D}, we can easily have
\begin{align}
\label{eqn_SINR_FD_U_decompose}
&\mathcal{C}_k^{1\text{,FD,U}} = \int\limits_{0}^{\infty} \mathbb{P} \left( \theta_k^{1\text{,FD,U}}(r) > T | D_k^{\text{FD}} = r \right) f_{D_{k}^{\text{FD}}}(r) dr,\\
\label{eqn_conditional_theta_k_FD_U}
&\mathbb{P} \left( \theta_k^{1\text{,FD,U}}(r) > T | D_k^{\text{FD}} = r \right) \nonumber\\
&= \exp\left( - \frac{T}{SSINR_{u,k}} \right) \prod\limits_{t = k, \bar{k}} \mathbb{E}_{I_{t,k}^{1\text{,U}}} \left[ \exp\left( - \frac{T r^{\alpha}}{P_u} I_{t,k}^{1\text{,U}} \right) \right].
\end{align}
Using the interference expression \eqref{eqn_channel_1_uplink_interference}, we have
\begin{align}
\label{eqn_interference_U_expectation}
&\mathbb{E}_{I_{t,k}^{1\text{,U}}} \Big[\exp(-\frac{T r^{\alpha}}{P_u} I_{t,k}^{1\text{,U}})\Big] = \mathcal{L}_{\Phi_t^{\text{FD}}}(r, T) \mathcal{L}_{\Phi_{u,t}^{\text{FD}}}(r, T) \mathcal{L}_{\Phi_t^{\text{HD}}}(r, T),
\end{align} 

\begin{figure}[t]
	\centering
	\includegraphics[width=85mm]{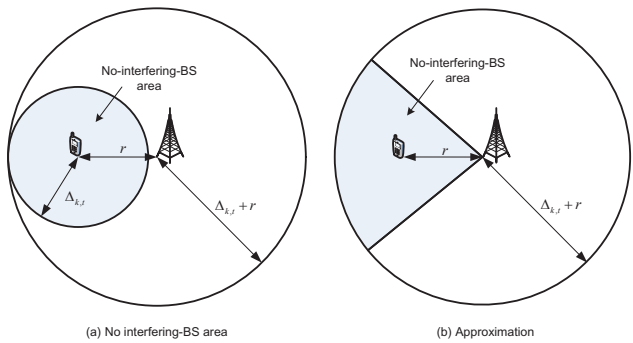}
	\vspace{-0.1cm}
	\caption{The no-interfering-BS area and its approximation.}
	\label{fig_No_interfering_BS_area}
	\vspace{-0.3cm}
\end{figure}

If the distance between the typical user and the tagged BS is $r$, there is no interfering BS in the shadowed area as shown in Fig. \ref{fig_No_interfering_BS_area} (a) on page $13$, considering the user association policy.
However, the no-interfering-BS area is difficult to characterize.
Hence, we make such an approximation that the no-interfering-BS area in Fig. \ref{fig_No_interfering_BS_area} (a) is approximated to the shadowed sector in Fig. \ref{fig_No_interfering_BS_area} (b).
The sector has the same area as the shadowed circular area in the left figure.
Thus we can have the interfering area ratio 
\begin{align}
p_{k,t}^I = 1 - \frac{\pi (\Delta_{k,t}(r))^2}{\pi (\Delta_{k,t}(r) + r)^2} \overset{(a)}{=} 1 - \left( \frac{P_t^{1 / \alpha}}{P_t^{1 / \alpha} + P_k^{1 / \alpha}} \right)^2.
\end{align}
Hence we have \eqref{eqn_Laplace_BS_FD_U} and \eqref{eqn_Laplace_BS_HD_U}.

\bibliographystyle{IEEEtran}
\bibliography{citation}
\newpage

\end{document}